\documentclass[10pt,letterpaper,twocolumn]{article}
\usepackage[T1]{fontenc}

\usepackage[letterpaper, width=7.00in, height=9.0in, columnsep=0.33in, top=1.0in, centering]{geometry}

\usepackage{graphicx} 
\usepackage{subcaption}
\usepackage{endnotes}
\usepackage{comment}
\usepackage{booktabs}
\usepackage{tabularx}
\usepackage{makecell}
\usepackage{array}
\usepackage{algorithm}
\usepackage{algpseudocode}
\usepackage{amssymb}
\usepackage{amsmath}
\usepackage{multirow}
\usepackage{xurl}
\usepackage{enumitem}
\usepackage{threeparttable}

\usepackage{authblk}

\usepackage{mathptmx} 

\usepackage[colorlinks=true, linkcolor=blue, citecolor=blue, urlcolor=blue]{hyperref}

\begin{document}

\title{\Large \bf Reality Check for Tor Website Fingerprinting in the Open World}

\author[1]{Mohammadhamed Shadbeh}
\author[1]{Khashayar Khajavi}
\author[1]{Tao Wang}

\affil[1]{Simon Fraser University}
\affil[ ]{\texttt{\{msa360, kka151, taowang\}@sfu.ca}}

\date{} 

\maketitle

\begin{abstract}

Website fingerprinting (WF) attacks on Tor can infer user destinations from encrypted traffic metadata. However, their real-world effectiveness remains debated due to laboratory settings that fail to capture network fluctuations, evaluate noise, and create a representative open world. In this work, we re-examine WF from a guard-relay vantage point using a novel, privacy-preserving methodology that builds an open-world background from real, unlabeled Tor traffic paired with synthetic monitored traces. Using this methodology, we collect a large-scale dataset of over 800,000 traces. We then benchmark state-of-the-art WF attacks under a cross-network setting and show that WF remains highly effective against real Tor open-world traffic: the best-performing attack achieves 0.956 precision and 0.922 recall at a 9\% base rate. We further present results that demonstrate robustness to small training sets, network jitter, and concept drift. Moreover, we show that timing-independent classifiers are significantly more robust to network variability than others. Finally, we provide the first systematic study of Tor's Conflux traffic-splitting, where we show that a guard node with a latency advantage can maintain high attack effectiveness even when traffic is split.

\end{abstract}

\section{Introduction}

Tor is the most widely deployed anonymity network for web browsing, designed to protect users against network surveillance~\cite{dingledine2004tor}. To access an Internet resource, a Tor client constructs a circuit through three relays (or nodes): a guard, a middle, and an exit, all operated largely by volunteers. Traffic traverses this circuit protected by layered encryption, ensuring that no single relay can simultaneously learn both the user's network identity and the destination being accessed.

A central privacy goal of Tor is to prevent \emph{linkability} between a user and the website they visit~\cite{cherubin2022online,jansen2023repositioning}. When an adversary can infer a user’s destination, this goal is violated. Website fingerprinting (WF) attacks aim to establish such linkability by passively analyzing encrypted Tor traffic and predicting the visited website or webpage from observable metadata~\cite{wang2013improved,wang2014effective,wang2016realistically}. Over the years, much research has been conducted to test, verify, or even break the Tor network anonymization assumption~\cite{hayes2016k,sirinam2018deep,Rahman_2020,shen2023subverting,deng2024robust}. This body of research targets a fundamental vulnerability in such a system: while the encrypted traffic is designed to anonymize the data and hide the content of data transfer, it cannot fully obscure the patterns of data such as packet timing and direction. Research has shown that this \textit{leaked} data is sufficient for linkability.

However, translating high laboratory accuracy into reliable real-world de-anonymization remains a challenging task~\cite{perry2013critique,juarez2014critical,cherubin2022online}. Prior work has stressed that many WF evaluations implicitly rely on conditions that are hard to guarantee outside of controlled crawls, such as stable client settings and browser versions~\cite{juarez2014critical}, consistent network conditions (latency and bandwidth)~\cite{cherubin2022online}, clear page-load boundaries~\cite{wang2016realistically}, and largely sequential browsing without substantial background activity or overlapping tabs~\cite{guan2021bapm,jin2023transformer,deng2026towards}. Moreover, in realistic open-world settings the \emph{base rate} of monitored-page visits is typically low; thus, even modest false-positive rates can render an attack impractical~\cite{wang2020high}. Consistent with these concerns, recent work by Cherubin et al., measuring WF under more realistic conditions, argues that while WF may succeed in narrowly targeted scenarios, deploying effective attacks at scale remains difficult~\cite{cherubin2022online}. This gap between laboratory assumptions and real-world deployment motivates a careful re-examination of WF attacks under practical conditions.

\subsection{Our Contributions}

In this work, we develop a new methodology to analyze WF in realistic conditions by using a different vantage point than previous work: our adversary controls a Tor \emph{guard} relay and seeks to de-anonymize the clients that connect to it. We use real Tor users’ traces to build a large, real open world for WF analysis collected at the guard. Crucially, unlike prior work, this collection is strictly privacy-preserving: we cannot obtain any ground truth or make any attempt to classify such traffic. In addition, we complement this real-world background traffic with synthetic monitored traces collected at the guard, and without losing our experimental validity, we evaluate WF performance in realistic cross-network conditions.

We adopt this vantage point and explicitly analyze and justify why it is a particularly relevant and powerful setting for a realistic adversary. 
Our methodology combines (i) \emph{synthetic} monitored traces: collected by controlled clients in different geographical locations to account for evaluation bias (e.g., client/network heterogeneity), subjected to an extensive, novel sanitization pipeline, and (ii) \emph{real} open-world (non-monitored) traces: collected at the guard from genuine Tor traffic.

Our open-world evaluations show that modern WF attacks remain highly effective against real-world traffic, even under rigorous cross-network conditions. We benchmark state-of-the-art classifiers, specifically analyzing their robustness to network variability and the impact of timing-dependent features on performance. Finally, we provide the first systematic study of Tor’s traffic-splitting mechanism, Conflux~\cite{alsabah2013path,tor-prop329}. We show that although Conflux distributes traffic across multiple circuits (or ``legs''), a resourceful guard adversary can still mount effective WF attacks.

In summary, our contributions are as follows:

\begin{itemize}[topsep=5pt]

    \item We introduce a rigorous and privacy-preserving guard-adversary methodology that captures a \emph{real} Tor open world without collecting IP addresses or destinations. This novel methodology removes the unrealistic advantages of the standard laboratory setting while avoiding the unnecessary disadvantages imposed by other real-world studies~\cite{cherubin2022online}. By instrumenting a guard relay to log per-cell metadata using ephemeral identifiers and strict sanitization, we construct a large-scale dataset of over 800,000 traces. We release both the dataset and our analysis code (see the \textit{Open Science} appendix).

    \item We analyze the effectiveness of Tor Guards as website fingerprinting attackers for the first time. We show that the Guard has unique advantages as an attacker: it is able to remove control data, identify individual circuits, and de-multiplex packet data from simultaneously loaded webpages. We evaluate these advantages and compare the results to a scenario where the attacker lacks access to this metadata. We show that we can successfully apply WF attacks in both scenarios.

    \item To our knowledge, we are the first to demonstrate that WF attacks achieve strong open-world performance against real-world Tor traffic. Under cross-network conditions, Deep Fingerprinting (DF)~\cite{sirinam2018deep} achieves an $r$-precision~\cite{wang2020high} with $r=10$ ($\pi_{10}$) of 0.956 and a recall of 0.922; pooled training/testing achieves $\pi_{10}={0.980}$ and recall of 0.968. Benchmarking state-of-the-art attacks also reveals that DF generalizes best under cross-network shifts, while Robust Fingerprinting (RF)~\cite{shen2023subverting} shows superior robustness to concept drift and traffic fragmentation.

    \item We provide the first systematic evaluation of WF under Tor's Conflux traffic-splitting. While single-leg observation causes performance to drop substantially, we show that a more powerful guard with a reasonable latency-advantage observing the primary leg recovers substantial attack effectiveness. Such a powerful guard results in an increase in recall from 0.522 to 0.881 (at FPR = 0.5\%).

\end{itemize}

\subsection{Paper Organization}

The remainder of the paper is as follows. Section~\ref{sec:related_work} reviews the background and related work, and Section~\ref{sec:methodology} details our guard-adversary methodology and data pipeline. Section~\ref{sec:experiments} presents our experimental results, Section~\ref{sec:discussion} discusses the implications. Finally, Section~\ref{sec:conclusion} concludes with future directions.

\begin{figure*}[t]
    \centering
    
            \begin{subfigure}[b]{0.48\textwidth}
        \centering
        \includegraphics[width=\linewidth]{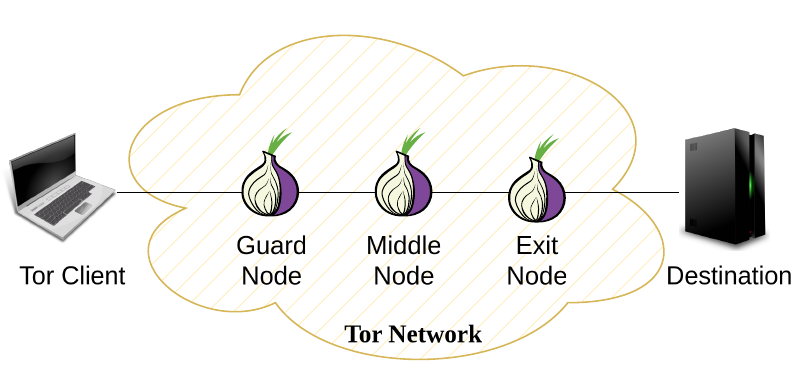}
        
                        \vspace{1.7em} 
        
        \caption{Standard Tor Circuit}
        \label{fig:tor-basics}
    \end{subfigure}
    \hfill
            \begin{subfigure}[b]{0.48\textwidth}
        \centering
        \includegraphics[width=\linewidth]{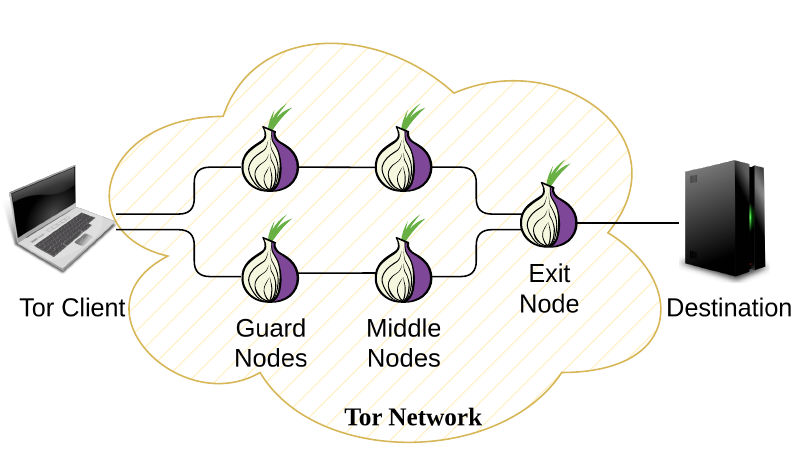}
        \caption{Tor Under Conflux}
        \label{fig:tor-basics-cfx}
    \end{subfigure}
    
        \caption{Overview of Tor traffic routing architectures. \textbf{(a)} In a standard Tor configuration, traffic flows sequentially through three intermediate nodes: (i) the guard node, which knows the client's IP address but not the destination; (ii) the middle node, which merely forwards the encrypted traffic; and (iii) the exit node, which connects to the destination server but remains oblivious to the client's original IP address. \textbf{(b)} Under Conflux, traffic between the client and the exit node is dynamically split across multiple linked legs. Each Conflux leg routes traffic similar to a standard Tor circuit. Both legs share the same exit node but use different guard and middle nodes.}
    \label{fig:tor_vs_conflux}
\end{figure*}

\section{Background \& Related Work}
\label{sec:related_work}

\subsection{Tor}

Tor~\cite{dingledine2004tor} is a low-latency anonymity network primarily used for web browsing. A Tor client accesses the Internet through a circuit that typically comprises three volunteer-operated relays (or nodes): an entry (guard), a middle, and an exit. Traffic is relayed as fixed-size cells ($\approx$512~bytes) and protected with layered encryption so that no single relay learns both the client’s network identity and the destination (Figure~\ref{fig:tor-basics}). In practice, most users access the network via Tor Browser, which enforces additional defenses against browser fingerprinting and tracking. When a circuit is established, both the ISP and the guard node can observe the user's real IP address, while the destination is hidden by design. Anonymity is compromised if an adversary can link these two pieces of information, motivating attackers who observe traffic near the client or control entry relays.

\subsection{Website Fingerprinting}
\begin{figure}[htbp]
    \centering
    \includegraphics[width=1\linewidth]{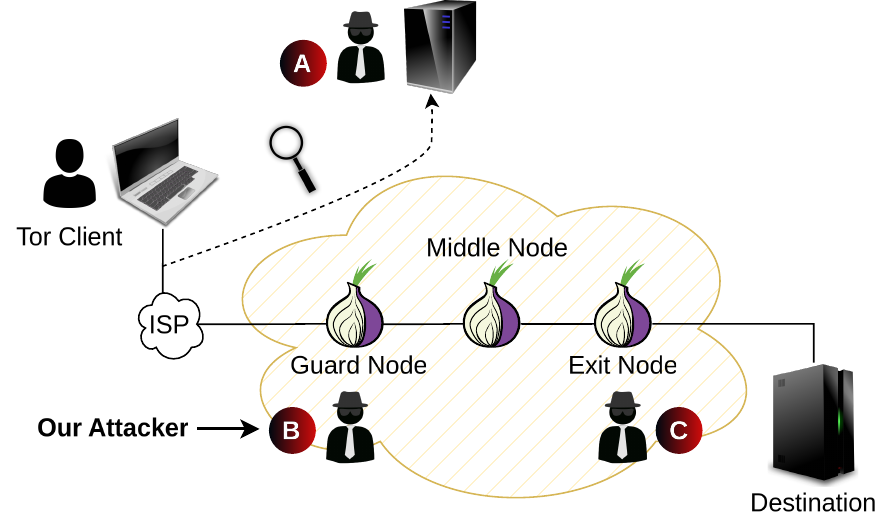}
    \caption{An attacker can place themselves in different points of entry in a Tor network to train their classifiers for WF attacks. Traditionally, the attacker is assumed to be an eavesdropper who is observing the traffic to/from clients (Attacker A). For this, traces are commonly gathered through the Tor client itself to train the classifiers. Our approach is to gather both training and testing data from a Tor guard node (Attacker B). Cherubin et al.~\cite{cherubin2022online} gather the training data through a Tor exit node (Attacker C). }
    \label{fig:tor-wf-guard-link}
\end{figure}

Website fingerprinting (WF) attacks aim to infer the website (or webpage) a user visits over Tor by passively analyzing encrypted traffic metadata, thus violating Tor’s goal of preventing destination-user linkability. WF is typically formulated as a supervised classification task: the adversary collects labeled training traces by repeatedly visiting target destinations with controlled clients and recording observable features such as cell directions and timings. Because a WF attacker is positioned on the client’s access link (e.g., the ISP) or acts as a malicious Tor guard relay, they know the client's network identity and seek to recover the hidden destination (Figure~\ref{fig:tor-wf-guard-link}).

WF is evaluated in two canonical settings: in the \emph{closed world}, the client visits only a fixed set of monitored destinations, reducing the attack to a multi-class classification task; in the \emph{open world}, the client may also visit arbitrary non-monitored destinations. In this setting, the attacker must detect whether a trace belongs to the monitored set (and if so, which site) or reject it as non-monitored. Open-world evaluations require analyzing precision and recall while accounting for the low base rate of monitored visits, and are widely viewed as a more realistic benchmark to assess practical risk~\cite{wang2020high,juarez2014critical}.

A large body of work demonstrates high WF performance in a \emph{standard laboratory setting}, where both monitored and non-monitored traces are collected at the client via browser automation under controlled conditions~\cite{panchenko2016website,sirinam2018deep,Bhat_2019}. While some studies explore more challenging regimes such as early classification, limited training data, and low base rates~\cite{deng2024robust,sirinam2019triplet,wang2016realistically}, others caution that laboratory assumptions can significantly inflate performance estimates. Specifically, laboratory traces are typically clean and \emph{page-load-aligned}, meaning the traffic capture strictly corresponds to the start and end of a single page load. Realistic browsing, by contrast, involves overlapping page loads (the ``multi-tab'' problem) and background activity; these factors blur trace boundaries and inject noise, often significantly degrading attack success~\cite{juarez2014critical,wang2016realistically}. Subsequent research has introduced methods to mitigate parts of these issues, demonstrating improved robustness against multi-tab browsing and background traffic~\cite{xu2018multi,automated_multitab_wf,robust_multitab_wf,wang2016realistically}.

Most closely related to our focus on \emph{realistic} WF, Cherubin et al.~\cite{cherubin2022online} conducted the first large-scale evaluation based on traffic from real Tor users. Their methodology trains a classifier using traffic collected at an exit relay (where the destination is visible) and evaluates its performance at a guard relay. They report that accuracy in this setting is substantially lower than in laboratory evaluations, finding the attack effective only for small monitored sets. Our work builds on these insights but revisits the training and evaluation methodology to demonstrate that WF remains a potent threat in practice.

\subsection{Circuits and Guards}
\label{sec:circuits_and_guards}

Tor clients communicate with their guards over long-lived TLS connections, referred to as \emph{channels}~\cite{tor-spec}. In the Tor protocol, a channel ID represents this single TLS connection between a client (or relay) and the guard. On the guard node, these IDs are assigned sequentially, starting from 0 upon the Tor process startup, and remain constant only for the duration of the TLS connection. Because these IDs are merely local counters that are reset upon restart, they possess no global significance and cannot be used to track users across different guards or different sessions.

Within a single channel, multiplexing allows for multiple concurrent circuits. To facilitate this, the Tor client generates a random 32-bit integer for each circuit it establishes. This circuit ID is included in the header of every cell sent to the guard. Since Tor clients do not persistently record these IDs and destroy them upon circuit closure, the circuit ID acts as a temporary, random tag. 

Despite the temporary nature of these routing identifiers, the guard remains a particularly dangerous vantage point for an adversary. Upon initialization, a client selects a small set of entry guards and retains them for an extended period (typically months). This persistence means a malicious or compromised guard can observe both the client’s IP address and their incoming traffic patterns over long durations. Consequently, we adopt the guard-level adversary model in this work.

Furthermore, because guards observe relay-level metadata (e.g., circuit IDs and protocol/control artifacts), they can leverage Tor's \emph{stream isolation} mechanism (Section~\ref{sec:stream_isolation}). Under stream isolation, requests initiated from different first-party contexts are assigned to distinct circuits (or, under Conflux, to distinct linked sets of legs). The guard can use the visible circuit IDs to effectively demultiplex concurrent page loads. Unlike local network adversaries (e.g., ISPs) who struggle to disentangle overlapping traffic (the ``multi-tab problem''~\cite{automated_multitab_wf}), a guard can separate target traffic from background noise, such as video streaming or parallel browsing, simply by filtering on the circuit ID.

\subsection{Stream Isolation}
\label{sec:stream_isolation}

Tor Browser enforces stream isolation: it uses different circuits for different first-party domains, determined by the address bar of the tab in Tor Browser~\cite{stream_isolation}. For example, if a user opens \textit{test.example.com} in a new tab, Tor creates a circuit for \textit{example.com}, and all resources loaded by that tab (regardless of their domains) are sent over the same circuit. If the user opens \textit{example.org} in another tab and loads its resources, the Tor process uses a different circuit for those requests.

Tor Browser sends its requests to the Tor process through a SOCKS5 proxy. The proxy’s username and password fields are used for Tor’s isolation decisions. When a user opens a new webpage, Tor Browser generates a SOCKS5 request and sends it to the Tor process. In this request, Tor Browser sets the SOCKS5 username to the first-party domain of the tab from which the request originates and sets the SOCKS5 password to a unique nonce for that domain. We note that, as part of the same proxy request, Tor Browser also includes the domain name of the request itself. This domain name reflects the true destination of the request and is not necessarily the same as the proxy username, which reflects the first-party domain in the tab’s address bar.

For a request originating from Tor Browser to be considered part of the same stream, the proxy request must share the same username and password. For each new request, Tor compares the username and password against all active circuits; if no suitable circuit is found, it creates a new one. As a result, stream isolation confines the loading of a webpage to a single circuit.

The introduction of Conflux (Section~\ref{sec:conflux}) changes the above stream-isolation assumptions, as it is possible for two different circuits to be used to load a webpage. However, Tor still enforces stream isolation under Conflux: each time a new first-party domain needs to be reached, a new set of Conflux legs is created and linked together. Tor also tracks the number of actively linked Conflux sets. Thus, under Conflux, Tor stream isolation applies to a \emph{pair} of circuits rather than a single circuit.

\subsection{Conflux}
\label{sec:conflux}

Tor introduced Conflux~\cite{tor-prop329} in the official stable Tor release v0.4.8.4. Conflux is a multipath traffic-splitting mechanism designed to reduce congestion and head-of-line blocking by distributing a single traffic stream across multiple circuits. Conflux is negotiated end-to-end and is enabled only when both the client and the exit relay support the protocol. When this protocol is active, the client establishes two (or more) independent circuits, referred to as \emph{legs}, which share the same exit relay but typically traverse different guards and middle relays.\footnote{While Conflux generalizes to more than two legs, Tor uses two by default; we therefore focus on the two-leg case.} The client and exit perform a handshake to link these legs into a single logical connection, allowing cells to be scheduled dynamically over either leg (see Figure~\ref{fig:tor-basics-cfx}).

Tor designates one leg as \emph{primary} and the other as \emph{secondary}; importantly, the client and exit make these decisions independently and may switch the primary leg dynamically. Under the default \emph{LowRTT} scheduling strategy, the primary leg is defined as the circuit with the lowest current RTT that has available space in its congestion window~\cite{tor-prop329}. Tor measures RTT during the initial handshake and continuously updates it as traffic flows, dynamically deciding whether to retain or switch the primary leg. Consequently, this continuous evaluation can lead to frequent switching between legs as conditions change. Conflux remains backward compatible; if a second leg cannot be created or linked (e.g., due to resource limits), traffic proceeds over a single circuit as in standard Tor.

From a WF perspective, Conflux does not alter the amount of data accessible to a local attacker (e.g., the ISP), as they observe the client’s full network link. A guard attacker, however, observes only the cells carried on its own leg rather than the complete traffic stream. In our work, we analyze Conflux under the default LowRTT configuration and quantify how traffic splitting impacts the efficacy of open-world WF attacks in the guard-adversary setting.

\section{Methodology}
\label{sec:methodology}

In this section, we describe our novel data collection methodology for evaluating website fingerprinting. We begin by explaining the key insight that allows us to collect real Tor traffic for WF without compromising privacy. We then detail our data collection pipeline for monitored and non-monitored traffic, followed by the sanitization steps used to extract clean, page-load-aligned traces from the raw data. Next, we summarize our technical modifications to instrument Tor components for collection and ground truth. Finally, we detail the inherent privacy-preserving measures of this pipeline.

\subsection{Description of Our Methodology}
\label{sec:adversary-model}

\begin{table}[t!]
  \centering
  \caption{Comparison of experimental designs across settings.}
  \label{tab:cherubin_vs_us}
  \small
  \renewcommand{\arraystretch}{1.2}
  
    \begin{threeparttable}
    \begin{tabularx}{\linewidth}{@{} l >{\arraybackslash}X >{\arraybackslash}X >{\arraybackslash}X @{}}
      \toprule
      \textbf{Setting} & \textbf{Standard} & \textbf{Cherubin} & \textbf{Ours} \\
      \midrule
      \multicolumn{4}{@{}l}{\textsc{Training Set}} \\
      \midrule
      Monitored traffic     & Synthetic & Real    & Synthetic \\
      Non-monitored traffic & Synthetic & Real    & Real \\
      Vantage point         & Client    & Exit    & Guard \\
      Class label           & Webpages  & Domains & Webpages \\
      \midrule
      \multicolumn{4}{@{}l}{\textsc{Testing Set}} \\
      \midrule
      Monitored traffic     & Synthetic & Synthetic & Synthetic \\
      Non-monitored traffic & Synthetic & Real      & Real \\
      Vantage point         & Client    & Guard     & Guard \\
            Class label           & Webpages  & Domains\tnote{*}  & Webpages \\
      \bottomrule
    \end{tabularx}
    
        \begin{tablenotes}
      \item[*] They crawl a set of webpages from which domain labels are derived at the exit. 
    \end{tablenotes}
  \end{threeparttable}
\end{table}

We explain our new methodology via Table~\ref{tab:cherubin_vs_us} by focusing on the improvements made by Cherubin et al.~\cite{cherubin2022online} (hereafter the ``Cherubin setting'') over the standard laboratory setting, and how we improve over the Cherubin setting. 

In the standard setting, both monitored and non-monitored traces are synthetically generated via automated browser crawls at the client. Synthetic monitored traces effectively model a low-noise client visiting target pages in a manner that yields clean, page-load-aligned data. While generalizing this behavior to all real users requires caution, it is reasonable to argue that if specific browsing behaviors facilitate de-anonymization, WF poses a valid threat. However, using synthetic non-monitored traces introduces significant validity concerns: they fail to capture the complexities of the real Tor open world, as they lack realistic popularity distributions, browsing patterns, and the dynamically generated content that dominates real-world traffic.

The Cherubin setting mainly tackles the latter issue. In the Cherubin setting, non-monitored traces for both training and testing are real Tor traces collected at the exit relay (Figure~\ref{fig:tor-wf-guard-link}). Monitored traces for training are also collected at the exit relay. 
Notably for training, this prevents page-level labeling for ground truth: only the domain can be observed at the exit relay (via DNS). Consequently, all pages within a domain are collapsed into a single class even though different webpages have different traffic patterns; in the worst case, a classifier may never have the opportunity to train on the specific monitored webpage that appears at test time. For evaluation in the Cherubin setting, monitored test traces are generated synthetically via automated crawls and collected at the guard.

The key observation that drives our new methodology is that \emph{forcing an adversary to train on exit-collected monitored traces is an unnecessary handicap}, mainly because it confuses domain and page labels. 
A practical attacker can choose to train on synthetic or real traffic according to which would give them the best performance. 
Comparing the Cherubin setting with ours, the {\em only} difference is the attacker’s training strategy: 
our monitored training set is synthetic, allowing the attacker to label them correctly as the real webpage rather than the domain. 
Our open world (non-monitored set) still uses real Tor traffic.
Our testing task for the attacker is exactly the same as the Cherubin setting except for the label: the classifier attempts to identify synthetic monitored traces as well as real non-monitored traces.

Training on monitored traces collected at the exit and testing on those collected at the guard also creates an unnecessary obstacle for the attacker in the Cherubin setting. 
Even if the underlying page load is the same, these traces differ due to relaying effects that alter timing and packet order, leading to a train-test mismatch that punishes the classifier~\cite{jansen2023repositioning}. 
By aligning the training and evaluation vantage points at the guard, our methodology eliminates this mismatch, providing a more accurate model of a real-world adversary.

Finally, because our setting permits the adversary to train on synthetically generated and precisely labeled monitored traffic, we must ensure this does not artificially inflate attack performance. To this end, we (i) generate training and testing traces using clients on distinct networks with substantially different latency profiles relative to the guard, (ii) temporally separate monitored visits across these clients to prevent the models from learning transient network conditions, and (iii) carefully trim each trace to eliminate non-page-load artifacts at the boundaries. We next describe our data-collection methodology in detail.

\subsection{Data Collection}

Table~\ref{tab:overall-traces} details the specific datasets used for the experiments in this work, representing the high-quality subset retained after the sanitization process (Section~\ref{sec:data-sanitization}). We conducted data collection in two distinct phases: \emph{pre-Conflux} and \emph{post-Conflux}, corresponding to periods before and after the protocol's full deployment on the Tor network. In both phases, we selected monitored webpages from sites ranked outside the top 10{,}000 in the Tranco list (i.e., rank~$> 10{,}000$) at the time of collection~\cite{LePochat2019}. This selection strategy is justified in Section~\ref{sec:non-mon-traffic}. For the open-world (non-monitored) component, pre-Conflux traces were drawn from guard observations prior to the protocol's introduction, whereas post-Conflux traces were collected after its widespread adoption.

\begin{table}[t]
\centering
\caption{Summary of sanitized monitored and non-monitored traces used in our experiments for the pre- and post-Conflux datasets. Monitored traces were collected via our clients in Canada (CA), Australia (AU), and the United Kingdom (UK). We categorize traces into three types: normal traces (N), traces with added latency (L), and traces gathered through the clients (C) as opposed to the guard node.}
\label{tab:overall-traces}

\normalsize
\renewcommand{\arraystretch}{1.5}

\resizebox{\columnwidth}{!}{%
    \begin{tabular}{l c c c c c}
    \toprule
    \multirow{2}{*}{\shortstack{Dataset\\(\# Classes)}} & \multirow{2}{*}{Client} & \multicolumn{3}{c}{Monitored Traces} & \multirow{2}{*}{\shortstack{Non-\\Mon.}} \\
    \cmidrule(lr){3-5}
     & & N & L & C & \\
    \midrule
    
        \multirow{3}{*}{\shortstack{Pre-Conflux\\(103)}} 
        & CA & 21,068 & - & - & \multirow{3}{*}{79,833} \\
        & AU & 20,660 & - & - & \\
        & UK & 61,641 & - & - & \\
    \cmidrule{1-6}
    
        \multirow{3}{*}{\shortstack{Post-Conflux\\(112)}} 
        & CA & 54,485 & 96,420 & 176,774 & \multirow{3}{*}{31,724} \\
        & AU & 61,272 & - & 71,834 & \\
        & UK & 42,476 & - & 86,920 & \\
    \bottomrule
    \end{tabular}%
} 
\end{table}

We emphasize that using synthetically generated monitored traces for testing aligns with prior realistic WF evaluations (e.g., Cherubin et al.~\cite{cherubin2022online}). In our methodology, these traces provide precise webpage-level ground truth, while their use in training enables the adversary to learn fine-grained traffic patterns without compromising the realism of the evaluation. We detail each phase of our approach below.

\subsubsection{Pre-Conflux Monitored Webpages}
\label{method:monitored}

For the pre-Conflux dataset, we selected 103 monitored webpages from sites ranked between 10{,}000 and 10{,}243 in the Tranco list, retaining only those that were consistently reachable during collection. We excluded pages that were inaccessible or repeatedly triggered CAPTCHA challenges over Tor, as they produce unstable and unrepresentative traces.

We collected monitored traces using controlled clients stationed in distinct network environments (detailed in Section~\ref{sec:setup}). To mitigate temporal bias and the influence of transient network conditions, the clients visited the target webpages in a round-robin schedule, adhering to standard WF data collection practices~\cite{wang2014effective,sirinam2018deep}. In total, each client visited each webpage at least 200 times to provide a rich dataset for evaluation. Primary data collection occurred at our guard relay, which identified traffic from our controlled clients via a real-time filter on their source addresses. Crucially, this was done without storing any IP addresses in the dataset. 

\subsubsection{Post-Conflux Monitored Webpages}
\label{sec:data-gathering-cfx}

To evaluate the impact of Conflux on fingerprintability, we collected a dedicated post-Conflux dataset. Given the temporal gap between measurement phases, we refreshed our monitored webpage list using an updated Tranco snapshot, while maintaining our selection criterion of sites ranked outside the top 10{,}000. In total, we selected 112 monitored webpages from sites ranked between 10{,}000 and 10{,}300.

We used a procedure similar to the one described in Section~\ref{method:monitored} to collect post-Conflux monitored traces using our controlled clients. However, analyzing Conflux traces collected through a guard node breaks the conventional WF assumption of having access to the full trace. Therefore, to ensure the highest quality for our Conflux evaluation, each of our crawlers visited each webpage at least 400 times. Furthermore, collecting Conflux traffic from a guard vantage requires ensuring that at least one Conflux leg traverses our guard relay. To facilitate this, we modified the guard selection logic of our controlled Tor clients during Conflux data collection. Following established measurement methodologies, we disabled \texttt{UseEntryGuards} to bypass standard long-term guard persistence~\cite{Rimmer_2018}. We then configured the client to pin exactly one leg to our guard, while allowing the second leg to be selected from the public consensus using Tor’s default weighted selection algorithm. This configuration ensures guard-side visibility while preserving the realistic traffic splitting and scheduling dynamics of the Conflux protocol.

\subsubsection{Data Collection at the Client}
\label{sec:data-collection-client}

To facilitate accurate ground truth labeling, we logged lightweight metadata at our clients. For each page visit, the client recorded: (i) the first-party domain of the active tab; (ii) the request timestamp; (iii) the target domain (which may differ from the tab's first-party domain); and (iv) the circuit identifier. We utilized this metadata to reliably correlate each monitored page load with its corresponding guard-observed trace.

For the post-Conflux dataset, we further instrumented the client to log granular Conflux metadata for each page load, including the operational status of the Conflux set, the circuit identifiers of the constituent legs, and the RSA identity digests of each leg's guard. This metadata is crucial for accurately correlating client-side logs with traces captured at our guard relay.

We also recorded an additional subset of post-Conflux traces directly at the clients. Unlike the guard node, the client can decrypt Tor cells, enabling us to identify the specific cell types. We leveraged this decrypted data as ground truth for our guard-side First-Segment detector (Section~\ref{sec:fs-detection}). We additionally used this data to quantify Conflux leg-switching behavior.

\subsubsection{Non-Monitored Traffic}
\label{sec:non-mon-traffic}

To construct a realistic open-world background, we recorded guard-side traces from clients distinct from our controlled crawlers. Concretely, any client--guard traffic that did not match our controlled-client filter was designated as \emph{non-monitored} and retained as an unlabeled trace. We only recorded privacy-preserved per-cell metadata (as detailed in Section~\ref{sec:guard-mod}). Importantly, we neither know nor attempt to infer the destination of non-monitored traces, and we strictly do not retain client IP addresses. A pivotal insight of our methodology is collecting a realistic open-world dataset, which allows for rigorous evaluation without compromising user privacy.

Since a relay may also function as a \emph{middle} hop (regardless of its guard status), we must further exclude traffic where our node is not acting as the entry guard. To achieve this, we leveraged Tor’s requirement for public relays to prove their identities during channel establishment using \texttt{AUTHENTICATE} cells~\cite{tor-spec}. This mechanism is intrinsic to the Tor protocol: if the connection initiator is a public relay, it \emph{must} authenticate, whereas standard clients do not. Consequently, authenticated channels carry relay-to-relay traffic and should be filtered out from our datasets. We note that traffic originating from unlisted relays (e.g., bridges) for which we are the middle node may be incorrectly categorized as client traffic, since such relays do not authenticate.

A remaining source of label noise is unavoidable. Since non-monitored traces are unlabeled, a client not under our control might coincidentally visit one of our monitored webpages. Such an instance would be incorrectly labeled as non-monitored, leading to an inflated false-positive rate and, consequently, a conservative evaluation of the attack's performance. \emph{We selected monitored pages from sites ranked outside the top 10{,}000 specifically to minimize this risk.} Assuming this ranking correctly represents Tor usage~\cite{mani2018understanding}, the probability that a random visit hits any one of 100 monitored pages (ranked outside the top 10{,}000) is at most $1/10{,}000$; in practice it is likely much lower due to the heavy-tailed (Zipf-like) distribution of web popularity and also because popularity rankings apply to domains rather than individual webpages~\cite{breslau1999web}. Finally, distinct characteristics of Tor traffic, including a heavier long tail and the usage of onion services, further reduce the likelihood that an arbitrary Tor user visit matches one of our specific monitored webpages~\cite{mani2018understanding}.

\paragraph{Pre-Conflux Non-Monitored Traffic}
Our guard relay served a massive volume of daily connections, of which only a tiny fraction yielded viable open-world webpage traces following data sanitization (Section~\ref{sec:data-sanitization}). For the pre-Conflux dataset, the guard averaged more than 30,000 Tor TLS connections (representing distinct users) and 1.3 million client circuits per day. By comparison, a relay-to-relay TLS connection is long-lasting, and as a result, such traffic accounted for fewer than 5,000 connections over a three-day span, though these carried roughly 1 million circuits per day. Ultimately, after data sanitization, we extracted high-quality, real non-monitored traces from about 1\% of the total circuits observed.

\paragraph{Post-Conflux Non-Monitored Traffic}
The post-Conflux non-monitored traces were gathered approximately eight months after the release of Conflux in Tor Browser~13.0, ensuring the dataset reflects a network environment where Conflux adoption is prevalent. Due to the nature of Conflux and its requirement for clients to establish additional circuits, the number of non-monitored circuits was significantly larger in our post-Conflux dataset. The guard served over 120,000 client TLS connections and 4.3 million client circuits per day (exceeding 8 million when including middle-node circuits). In this dataset, after removing the spam circuits, approximately 74\% of the remaining circuits were identified as non-Conflux circuits and excluded from the dataset (see Section~\ref{sec:circuit-removal}). This absence of Conflux can be attributed to several factors: older Tor clients that do not support the protocol, clients that disabled Conflux after reaching the maximum allowed number of active linked sets, or clients accessing onion services (which Conflux does not currently support). Ultimately, after data sanitization, approximately 0.1\% of all captured non-monitored post-Conflux traces were deemed viable for training our models.

\subsection{Technical Modifications}
\label{sec:tech-mods}

To support guard-side collection and Conflux-aware analysis, we made lightweight modifications to Tor components (guard relay, client, and Tor Browser) to expose the minimal metadata required for trace extraction and labeling. We detail the full technical implementation below.

\subsubsection{Guard Instrumentation}
\label{sec:guard-mod}

We instrumented a Tor guard relay with a lightweight patch ($\approx$300 LoC) to record per-cell metadata as traffic traverses the node, capturing circuit and channel identifiers, high-precision local timestamps, and directionality (incoming/outgoing). As detailed in Section~\ref{sec:non-mon-traffic}, the guard does not record this metadata for channels in which it is the middle node. Furthermore, following the standard conventions~\cite{sirinam2018deep}, we define directionality from the client's perspective: \textit{outgoing} (+1) refers to cells sent by the client (towards the Guard), and \textit{incoming} (-1) refers to cells received by the client (from the Guard).

To distinguish our controlled crawlers from background traffic, we provided the guard with a dynamic list of our clients' IP addresses. The guard used this list to segregate monitored and non-monitored traces into separate files in real-time: one file per controlled client and one file for all non-monitored data. To minimize downtime---which could cause our node to lose its Guard status---this IP list could be adjusted on the fly and applied via a \texttt{SIGHUP} signal without restarting the Tor software. Additionally, this list was used to exempt our controlled clients from the relay's built-in Denial-of-Service (DoS) defenses, preventing standard rate-limiting mechanisms from throttling our crawlers during the rapid circuit creation resulting from the continuous loading of webpages.

\subsubsection{Client Instrumentation}
\label{sec:client-mod}

We modified the Tor client source code ($\approx$200 LoC) to log minimal ground-truth metadata for each page load (as detailed in Sections~\ref{sec:data-gathering-cfx} and \ref{sec:data-collection-client} ). Leveraging stream isolation (Section~\ref{sec:stream_isolation}), we identified the first-party domain of the requested webpage by inspecting the SOCKS5 authentication credentials sent by Tor Browser. Correspondingly, the specific destination domain was retrieved directly from the SOCKS5 destination address field.

In addition to the Conflux modifications detailed in Section~\ref{sec:data-gathering-cfx}, we also adjusted specific default values in our Tor clients to facilitate data gathering for our post-Conflux monitored dataset. The default hard-coded maximum number of unused open circuits (14) was increased to 50. The maximum number of Conflux prebuilt sets (default 3) was increased to 20. The maximum number of Conflux linked sets was set to 60 rather than the default of 10. In all cases, these adjustments ensured that there were always ready-to-use Conflux sets available for our post-Conflux data gathering. Without these changes, because our crawlers continuously opened new webpages, our Tor clients would soon exhaust the available Conflux sets and revert to loading webpages using single circuits.

\paragraph{Tor Browser.}
To generate monitored traces, we automated page loads directly within Tor Browser using Tampermonkey, a lightweight in-browser scripting framework. We avoided Selenium-based automation, as prior work has shown systematic differences in how web servers respond to Selenium-driven browsers compared to webpage requests initiated by real users~\cite{cassel2022omnicrawl,krumnow2022how}. Furthermore, instead of relying on the browser's internal Tor Launcher extension, we used the official method to launch the modified Tor process as an external instance~\cite{start_tor_browser_gitlab}. We carefully replicated the official Tor Launcher's runtime configuration to ensure fidelity, particularly regarding stream isolation.\footnote{This ensures strict adherence to Tor Browser’s isolation logic, such as the \texttt{KeepAliveIsolateSOCKSAuth} flag.} Finally, to reduce potential network noise unrelated to page loads, we disabled automatic browser and extension updates during the collection period.

\subsection{Data Sanitization}
\label{sec:data-sanitization}

Raw guard-side traffic contains significant activity unrelated to singular, clean webpage loads, such as non-web traffic, relay-to-relay connections, and abusive connections (spam), and thus requires rigorous processing before it is suitable for WF evaluation. To address this, we implemented a dedicated sanitization pipeline designed to extract high-quality traces from the raw guard observations. 

Our data sanitization pipeline was strictly governed by two core principles: first, we did not perform any sanitization that a real-world adversary could not perform; and second, we aimed to eliminate any artifacts resulting from our experimental setting that could artificially exaggerate WF performance. We detail the steps of this methodology below.

\subsubsection{Main Circuit Selection}
\label{sec:main-circuit-selection}

Sanitizing the monitored traffic from our crawlers required identifying the \textit{main circuit} for each webpage. When obtaining circuit-based traces for our pre-Conflux set, a webpage is ideally loaded over a single circuit due to stream isolation (Section~\ref{sec:stream_isolation}); however, in practice it may use multiple circuits. We highlight three main reasons for this behavior and describe how we selected the main circuit:

First, if the client repeatedly fails to load a resource, it may abandon the current circuit and use a new one for the remaining requests for that first-party domain, in order to avoid page-load failure due to relay issues. If multiple circuits to the same first-party domain occurred in succession, we designated the circuit with the highest cell count as the main circuit.

Second, a new circuit is created when the first-party domain in the address bar of an open tab changes. Such a change may be triggered by HTTP, HTML, or JavaScript redirection. When this occurs, Tor Browser’s SOCKS5 credentials change, and stream isolation (Section~\ref{sec:stream_isolation}) mandates a new circuit for the remaining requests from that tab. In these cases, we ignored the resulting circuits because their first-party domains no longer matched the initial domains selected from the Tranco database.

Third, a new circuit can be created to support Opportunistic Onions, which allows a non-onion webpage to advertise an alternative onion service address. In this case, the client creates a new circuit to connect to the onion service. This has become a common practice following the introduction of Cloudflare Onion Services~\cite{sayrafi_2018,cloudflare_docs_2023}. According to W3Techs, as of January 2026, 21.2\% of all websites use Cloudflare CDN as their reverse proxy~\cite{w3techs-cloudflare-reverse-proxy-2026}. Opportunistic Onions introduce a new challenge for data sanitization. This mechanism leverages HTTP Alternative Services (Alt-Svc)~\cite{rfc7838}; however, a client can only receive Alt-Svc headers after a connection has already been successfully established. Even after receiving an Alt-Svc header, Tor Browser requires a few seconds to establish a new circuit to the onion service and verify its certificate. During this time, the webpage continues loading over the original (main) circuit. Previous research shows that the most crucial part of a webpage for successful fingerprinting is its initial load~\cite{deng2024robust}. As such, we could safely ignore the circuits created as a result of Alt-Svc headers. We identified these circuits by selecting those whose SOCKS5 username matched the correct first-party domain but whose requested domain name was an \textit{.onion} service.

We note that for our post-Conflux set, rather than identifying the main circuit for each page load, we identified the \textit{main set}. Each Conflux set consists of two linked circuits; therefore, the same techniques described above were used to select the main Conflux set.

\subsubsection{Circuit Removal}
\label{sec:circuit-removal}

Circuit removal was a carefully designed, multi-step process in our data sanitization pipeline to remove low-quality traces from our dataset. Circuit removal took place in three steps:

\begin{enumerate}

\item \textbf{Spam:} For non-monitored connections, we noticed a large number of abusive circuits: a specific channel sometimes repeatedly creates many circuits. We marked a channel as spam and ignored all its circuits if it created more than 10,000 circuits during its lifetime. Our analysis shows that this rule affects only a few high-volume TLS channels, which create a large number of circuits carrying almost no data. For our pre-Conflux dataset, removing traces from these TLS channels resulted in the removal of approximately 8\% of non-monitored traces. This number is much higher for our post-Conflux dataset, in which spam channels account for approximately 75\% of our total non-monitored traffic. We note that marking a channel as a spam was an implementation decision to facilitate faster data sanitization. Even without removing such channels, almost all their circuits would still be marked for removal due to containing a small number of cells (as detailed in Step 3).

\item \textbf{Handshake Validation} 

For pre-Conflux data, the first and second cells of a circuit must be outgoing and incoming, respectively ([+1, -1])~\cite{tor-spec}. These are the circuit’s handshakes that the guard captures. Furthermore, the immediate cell after the handshake must be outgoing ([+1]), as it is always the client who initiates the requests. As a result, if an incorrect cell start sequence in a circuit is detected, we ignored that circuit. This process removed an additional 5\% of our remaining non-monitored traces.

Similarly, for the post-Conflux dataset, each leg of a Conflux set when observed through a guard node must have the handshake sequence of ([+1, -1, +1, -1, +1])~\cite{tor-spec}. This data sanitization resulted in removing approximately 67\% of non-monitored traces that were not previously marked as spam. However, this pattern is a general pattern that also applies to many instances of non-Conflux circuits. To further remove circuits that are not Conflux, we look at the time gap between the 2nd and 3rd handshake cells and compare them to the time gap between the 4th and 5th handshake cells. As the client receives the corresponding handshake cell from relays, we expect the client to respond relatively fast, and as such, the difference between these two time gaps should not be large. On the other hand, a non-Conflux circuit handshake finishes after the second cell, and since commonly a circuit is created in advance before it is used, a large difference between these two time gaps is expected. Although this technique may not address all circumstances, we believe it provides an acceptable initial measurement for Conflux.\footnote{Testing this heuristic approach on the \textit{monitored} dataset for both pre-Conflux and post-Conflux achieved more than 99.99\% accuracy.} This rule resulted in removing an additional 22\% of the remaining post-Conflux non-monitored traces.

\item \textbf{Small Circuits:} We did not include traces that contained fewer than 200 cells (around 100~KB of data). If a circuit's length was less than 200 cells, we considered it to be too small for our experiment. Since crawling the smallest webpage in our monitored set results in a trace of around 300 cells, the existence of a smaller trace in our monitored dataset could be the result of different factors: (i) circuits with very few cells could correspond to unused circuits, which were created by the Tor client but never used, or (ii) they may be the result of a failure to load a webpage (e.g., timeout or triggered a CAPTCHA). In our non-monitored set, the existence of such a circuit, in addition to the above reasons, could also be the result of abusive traffic (spam). It may also be the result of loading a webpage smaller than the smallest webpage in our monitored set (in which case, it is guaranteed that it belongs to the non-monitored set and including such a circuit in our final dataset could artificially inflate the model's performance). As such, these traces can be discarded. Compared to the previous steps, this step is responsible for filtering out the majority of remaining low-quality non-monitored circuits. For the pre-Conflux dataset, this filter resulted in the removal of 99.4\% of the remaining pre-Conflux non-monitored set. Similarly, for the post-Conflux dataset, approximately 98.3\% of the remaining non-monitored circuits were removed.

\end{enumerate}

Finally, we emphasize that the above steps are primarily to extract high-quality traces from the raw non-monitored traffic before training the models to avoid unrealistic inflated results. Although these steps also apply to our monitored circuits, almost all circuits from these controlled clients satisfy the above requirements.

\subsubsection{Circuit’s Head Trimming}
\label{sec:head-trim}

We trimmed the start of each trace to eliminate protocol artifacts and measurement noise that would otherwise introduce an optimistic bias. During normal operation, the Tor client periodically creates new circuits and maintains them in a pool of available circuits so they can be assigned to upcoming streams. As a result, there can be idle time lapses on the order of minutes between the initial circuit creation and the actual start of application data transfer. Retaining this variable idle latency would create artificial timing features for classifiers.

To correctly capture the true start time of the data transfer, we stripped the initial circuit handshake observed at the guard. Specifically, we removed the first two cells for pre-Conflux circuits, and the first five cells for post-Conflux circuits, as these correspond purely to the handshake protocol.

\subsubsection{Tail Trimming}
\label{sec:tail-trim}

At the tail, we needed to eliminate potential patterns and optimistic biases introduced by the end of a page load. For example, to gather data using our crawlers, our automation script closed the Tor Browser 10 seconds after a full page load \textit{or} after a default timeout of 45 seconds. This artificial shutdown can create unintended patterns, as the browser may transmit final data while closing (e.g., JavaScript on a webpage may transmit additional data) or send TCP FIN/RST packets for open connections. Furthermore, in some cases, due to network issues or destination server configurations, the circuit may remain active even after the browser is closed.

To ensure unbiased model evaluation, we applied a rigorous, four-stage pruning process to both monitored as well as non-monitored traces:

\begin{enumerate}
    \item \textbf{Teardown Removal:} We strictly removed the last two cells of every circuit, as these are typically the result of the circuit's standard teardown process~\cite{stream_isolation}.
    
    \item \textbf{Gap-Based Pruning:} We dynamically identified and removed the tail if we detected an artificial gap caused by the browser shutdown. We searched for a time gap of at least 5 seconds between consecutive cells. If there were multiple time gaps in a circuit, we specifically isolated the last one. To qualify for this removal, the first cell immediately following this gap had to be an outgoing cell (since the client initiates the closing process). Additionally, the isolated tail had to either contain fewer than 100 cells, or its total duration had to be less than 1 second. These numbers were selected empirically to be a good compromise between removing too many cells and too few. This step effectively sanitized approximately 53\% of our monitored traces and 24\% of our non-monitored traces.
    
    \item \textbf{Duration Capping:} Our analysis showed that there were rare cases in which a circuit remained active even after the browser was closed. We suspect this may happen due to network issues or bad destination server setup. Therefore, we capped the maximum duration of all traces (both monitored and non-monitored) to the 99th percentile of the maximum circuit time observed in the monitored set for the respective experiment. This cap ranged from 42 to 47 seconds depending on the dataset. Introducing this cap ensures consistency by trimming circuits that hung open anomalously.
    
    \item \textbf{Length Truncation:} Finally, adhering to standard website fingerprinting conventions~\cite{sirinam2018deep}, we truncated all remaining traces to a maximum length of 5,000 cells.
\end{enumerate}

\subsubsection{Time-Based Segmentation}
\label{sec:time-based-seg}

To quantify the performance of WF attacks without the guard advantage (i.e., without access to circuit IDs for de-multiplexing), we also employed \emph{time-based segmentation} as part of our experiments. This approach simulates a local network adversary (e.g., an ISP) who can separate traffic by user (IP address) but cannot distinguish between concurrent circuits. This necessitated specific adjustments to the data sanitization pipeline.

\paragraph{Monitored Traces.}
For monitored data, we could not use the main-circuit selection process described in Section~\ref{sec:main-circuit-selection}, which assumes guard vantage as it relies on circuit IDs. Instead, at the client, we recorded the start and end timestamps of the page load. Then, we extracted all cells traversing the guard on the specific channel with this client during this interval. This resulted in traces that merge multiple overlapping circuits (including potential background noise). Furthermore, since circuits are typically pre-built before the page request, we could not filter for circuit handshakes (Section~\ref{sec:circuit-removal}) or apply circuit's head trimming (Section~\ref{sec:head-trim}). Instead, we aligned the trace start to the first \emph{outgoing} cell, as the client-initiated request begins with an outgoing cell. Finally, we applied the tail-removal heuristics detailed in Section~\ref{sec:tail-trim}.

\paragraph{Non-Monitored Traces.}
Segmenting discrete page-aligned traces from continuous, unlabeled traffic remains an open research challenge~\cite{panchenko2016website,juarez2014critical}. We approximated segmentation by grouping traffic according to Channel ID (analogous to an ISP separating users by IP address) and applying a greedy temporal clustering algorithm to segment the continuous stream:
\begin{enumerate}
    \item For a given channel, we identified all active circuits and sorted them by their start time.
    \item We iterated through the sorted list. For the first available circuit $C_{i}$, we defined a candidate trace window $[t_{start}, t_{end}]$ corresponding to the lifespan of $C_{i}$.
    \item We aggregated all cells on the channel that fall within this window into a single trace.
    \item To prevent duplication, we marked $C_{i}$ and all other circuits that overlap with this window as ``consumed'' and removed them from the list of candidate circuits.
    \item We repeated the process for the next available circuit in the list.
\end{enumerate}

Similar to the monitored set, we aligned each resulting trace to start with an outgoing cell and applied our standard tail-removal heuristics. This methodology allows us to simulate an adversary who cannot distinguish between circuit IDs but can reconstruct Tor cells from TLS records, a capability demonstrated in prior work~\cite{wang2013improved}.

Finally, we note that we explored time-based segmentation only for the pre-Conflux dataset. Our post-Conflux dataset has been deliberately gathered through the guard node to observe the effect of traffic-splitting on the models' performance. To an adversary who is observing the client's connections (as opposed to a guard adversary), Conflux does not provide any new challenges since such an adversary has access to the traffic of both Conflux legs.

\subsection{Privacy of Our Methodology}

Our data collection and processing pipeline adhered to strict privacy minimization principles, particularly regarding non-monitored traffic. Unlike exit-side labeling approaches, we strictly did not record real client IP addresses or destination addresses. Guard-side raw traces consisted solely of per-cell metadata (channel and circuit identifiers, directionality, and inter-arrival timestamps). We also normalized these traces by resetting the first cell’s timestamp to 0, thus removing absolute timing information. Channel and circuit IDs served only as ephemeral, relay-local tags for demultiplexing traffic (e.g., distinguishing circuits within shared TLS sessions) and facilitating trace extraction. Crucially, due to how these IDs are generated by Tor, they cannot be used to link back to real-world users.

As an additional safeguard, all raw logs were gathered and processed exclusively on secured, private servers following data protection best practices. For the publicly released dataset, we stripped all auxiliary metadata: non-monitored traces are shuffled and stored without channel or circuit identifiers, ensuring that individual traces are independent and unlinkable across sessions. Finally, we made no attempt to infer the destinations of non-monitored traces, treating them strictly as unlabeled background traffic. Further technical details on our data-handling practices and identifier stripping are provided in the \textit{Ethical Considerations} appendix.

\section{Experimental Results}
\label{sec:experiments}

In this section, we evaluate the realistic threat of state-of-the-art Website Fingerprinting attacks using our new data collection methodology. We begin with an overview of our experimental setup. Then, aligning with prior work, we assess performance on our pre-Conflux dataset, where the guard attacker retains visibility of full traces. Subsequently, we present findings from post-Conflux experiments. Finally, we simulate an adversary controlling a powerful guard node and compare those simulations against our empirical post-Conflux results.

\subsection{Experimental Setup}
\label{sec:setup}

\paragraph{Clients.} We deployed three controlled Tor clients in Canada, Australia, and the United Kingdom to generate monitored traces. The clients were placed in distinct network environments to induce heterogeneous latency and network conditions. The clients ran Debian~12 with a graphical desktop environment suitable for Tor Browser automation. The Canadian client used an Intel i7-6700 CPU with 32~GB RAM. The UK and Australia clients were hosted on the DigitalOcean platform (2~vCPUs, 4~GB RAM). The measured round-trip latency from each client to our guard was approximately 68~ms (Canada), 223~ms (Australia), and 79~ms (UK).

\paragraph{Guard Relay.} We operated a dedicated Tor guard relay hosted on an OVH server in Canada (Intel Xeon E3-1245v2, 32~GB RAM, $2\times480$~GB SSD) with a 100~Mbps uplink. To ensure stable selection as a guard by Tor clients, we ran this relay continuously for more than one year prior to data collection, allowing it to obtain and maintain guard status~\cite{tor_blog_guard}.

\paragraph{WF Attacks.} In this work, we evaluate five state-of-the-art Website Fingerprinting attacks: $k$-fingerprinting ($k$-FP)~\cite{hayes2016k}, Deep Fingerprinting (DF)~\cite{sirinam2018deep}, Tik-Tok~\cite{Rahman_2020}, Robust Fingerprinting (RF)~\cite{shen2023subverting}, and Holmes~\cite{deng2024robust}. $k$-FP employs a random forest classifier over a high-dimensional set of handcrafted traffic features. DF utilizes a CNN-based model trained on packet direction sequences (excluding timing), while Tik-Tok extends the DF architecture by incorporating timing information. RF aims to improve robustness by leveraging a Traffic Aggregation Matrix (TAM) representation with a CNN. Finally, Holmes uses a hybrid convolutional encoder trained via supervised contrastive learning to align partial traces with full-trace embeddings for early-stage fingerprinting. Unless stated otherwise, we adopt each method’s original training procedure and recommended hyperparameters; full settings are provided in Appendix~\ref{app:hyperparams}.

\subsubsection{Metrics}
\label{sec:metrics}

We report open-world performance using the notation and definitions of Wang~\cite{wang2020high}. Let $N_P$ and $N_N$ denote, respectively, the number of \emph{monitored} (positive) and \emph{non-monitored} (negative) test traces. Among monitored traces, we distinguish \emph{true positives} ($N_{TP}$, correctly identified monitored pages) from \emph{wrong positives} ($N_{WP}$, predicted as a monitored class but assigned to the wrong monitored class). Among non-monitored traces, \emph{false positives} ($N_{FP}$) are those incorrectly classified as monitored. We then compute the corresponding rates:
\[
TPR=\frac{N_{TP}}{N_P},\qquad
WPR=\frac{N_{WP}}{N_P},\qquad
FPR=\frac{N_{FP}}{N_N}.
\]
We refer to TPR as recall. 

To evaluate model performance consistently, whenever we show how a variable affects TPR, we fix FPR to 0.5\% for high precision. The rationale for this baseline, based on observed model behavior, is discussed in Section~\ref{sec:per-webpage-analysis-fnr}. We emphasize that a low FPR alone is insufficient for an effective attack; a useful open-world classifier must simultaneously maintain high recall (TPR) to ensure that improved precision is not achieved merely by rejecting the majority of monitored traces.

To account for the base-rate fallacy in open-world WF, we report \emph{$r$-precision}~\cite{wang2020high}, which makes the (unknown) real-world ratio between non-monitored and monitored visits explicit. Let $r$ denote the base ratio of negatives-to-positives in deployment. Then:
\[
\pi_r \;=\; \frac{TPR}{TPR+WPR+r\cdot FPR}
\]

Importantly, for correct calculation of $r$-precision, wrong positives are \emph{not} the same as false positives (they originate from monitored traces), and conflating them can significantly affect the results.

To summarize the trade-off between $r$-precision and recall into a single metric, we report the \emph{$F_1$ score}, calculated as the harmonic mean of $r$-precision and recall:
\[
F_1 \;=\; 2 \cdot \frac{\pi_r \cdot TPR}{\pi_r + TPR}
\]

We report $\pi_r$ (and the derived $F_1$) for representative values of $r$ specified in each experiment. To evaluate performance in a realistic open-world environment, we primarily conduct our experiments using $\pi_{10}$. This represents a scenario where the volume of unmonitored traffic is ten times greater than that of the monitored traffic. While a balanced scenario ($\pi_{1}$) is frequently used in the literature, we argue that $\pi_{10}$ provides a more discriminative benchmark by accounting for the precision decay inherent in larger unmonitored sets without reaching the noise-dominated regime of extreme ratios. A comprehensive sensitivity analysis across $\pi_{r}$ for $r \in \{1, 10, 100, 1000\}$ for all evaluated models is provided in Section~\ref{sec:r_sensitivity}.

Finally, following Wang~\cite{wang2020high}, when the number of false positives is small ($N_{FP}<10$), estimating FPR (and thus $\pi_r$) can become unstable. In this case, we upper-bound FPR using the Wilson score interval~\cite{brown2001interval} and propagate this bound through the definition of $\pi_r$, reporting a conservative lower bound on $r$-precision.

\subsection{Open-World Evaluation}
\label{sec:exp:ow-preconflux}

\begin{table*}[t]
\centering
\caption{Open-world WF performance at the guard using the pre-Conflux dataset. We report $r$-precision at $r{=}1$ ($\pi_1$) and $r{=}10$ ($\pi_{10}$), recall ($R$), and the $F_1$ score ($F_1$) calculated using $\pi_{10}$. We compare the baseline by Sirinam et al.~\cite{sirinam2018deep} against our traces under two regimes: cross-network (Train: AU, Test: CA) and pooled (Train \& Test: Both). For each model, the decision threshold is selected to maximize the $F_1$ score.}
\label{tab:ow-with-guard}

\normalsize

\renewcommand{\arraystretch}{1.2}

\begin{tabular}{@{}lcccc cccc cccc@{}} 
\toprule
\multirow{2}{*}{Classifier} & \multicolumn{4}{c}{Baseline (Sirinam et al.)} & \multicolumn{4}{c}{Train: AU / Test: CA} & \multicolumn{4}{c}{Train \& Test: Both} \\
\cmidrule(lr){2-5} \cmidrule(lr){6-9} \cmidrule(lr){10-13}
 & $\pi_1$ & $\pi_{10}$ & $R$ & $F_1$ & $\pi_1$ & $\pi_{10}$ & $R$ & $F_1$ & $\pi_1$ & $\pi_{10}$ & $R$ & $F_1$ \\
\midrule
$k$-FP (2016)    & 0.978 & 0.861 & 0.791 & 0.825 & 0.852 & 0.717 & 0.307 & 0.430 & 0.988 & 0.970 & 0.915 & 0.942 \\
DF (2018)      & 0.994 & 0.951 & 0.940 & 0.945 & 0.987 & \textbf{0.956} & \textbf{0.922} & \textbf{0.939} & 0.996 & 0.979 & 0.966 & 0.973 \\
Tik-Tok (2019) & 0.994 & 0.947 & 0.896 & 0.921 & 0.973 & 0.901 & 0.844 & 0.872 & 0.996 & 0.980 & 0.948 & 0.964 \\
RF (2023)      & 0.996 & \textbf{0.969} & \textbf{0.958} & \textbf{0.964} & 0.359 & 0.089 & 0.031 & 0.046 & 0.987 & \textbf{0.980} & \textbf{0.968} & \textbf{0.974} \\
Holmes (2024)  & 0.996 & 0.970 & 0.931 & 0.950 & 0.577 & 0.176 & 0.004 & 0.009 & 0.971 & 0.950 & 0.956 & 0.953 \\
\bottomrule
\end{tabular}
\end{table*}

\begin{figure*}[t]
    \centering
    \includegraphics[width=1\linewidth]{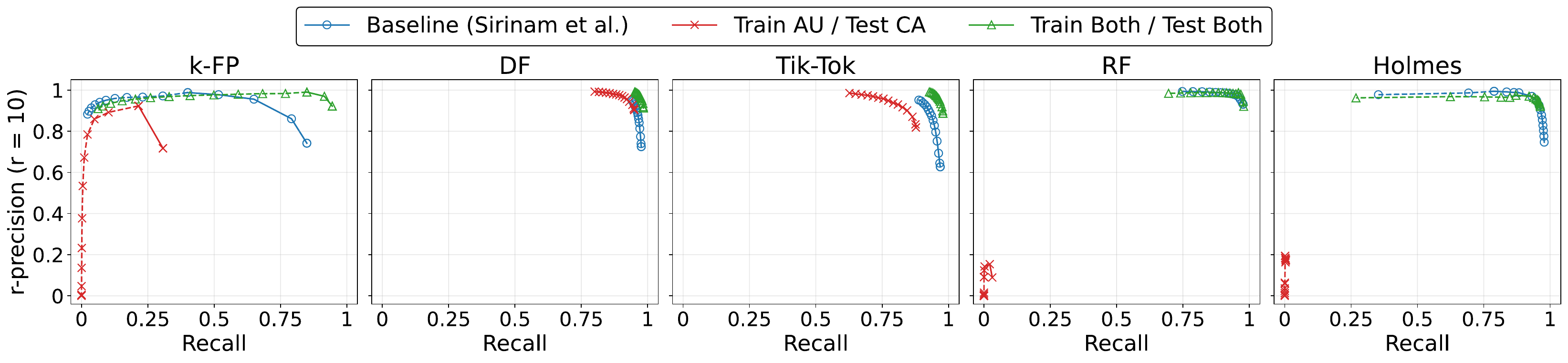}
    \caption{The $r$-precision ($r=10$) vs. recall curves for different classifiers across three scenarios of Table~\ref{tab:ow-with-guard}. Dashed lines (where applicable) represent $r$-precision corrections using the Wilson score interval when the number of false positives is below 10.}
    \label{fig:ow-combined-au-ca-breakdown-grid}
\end{figure*}

We begin by evaluating open-world website fingerprinting at the guard using our pre-Conflux dataset (Table~\ref{tab:overall-traces}) and the five representative attacks described in Section~\ref{sec:setup}. To include the impact of network and deployment mismatch, we evaluate under two regimes: (i) \emph{cross-network}, where we train on Australia (AU) monitored traces
and test on Canada (CA) traces;~\footnote{We verified that this setting is symmetric for the best attacks (i.e., train on CA, test on AU).} and (ii) \emph{pooled}, where CA and AU monitored traces are combined. 
We also include a time gap of one month in the real Tor non-monitored traces between the training and testing sets, similar to a real attacker who may face difficulties collecting such traces (we evaluate concept drift with monitored traces in Section~\ref{sec:concept-drift}).
In Table~\ref{tab:ow-with-guard}, we report recall ($R$) and $r$-precision for $r\!=\!1$ ($\pi_{1}$) and $r\!=\!10$ ($\pi_{10}$), along with the corresponding $F_1$ score (derived from $\pi_{10}$ and $R$). Since open-world performance depends on the decision threshold, for each method we select the operating point that maximizes this $F_1$ score. Finally, we benchmark performance against the widely used Sirinam et al.\ dataset~\cite{sirinam2018deep} (95 monitored websites with 1,000 traces per site and 40,000 open-world traces).

Table~\ref{tab:ow-with-guard} confirms that, on the standard laboratory setting (Sirinam et al.), all attacks achieve high recall and $r$-precision, with RF and Holmes slightly outperforming earlier methods, consistent with their design goal of leveraging richer representations beyond raw direction sequences. In contrast, performance diverges sharply in the cross-network setting (Train: AU, Test: CA). Here, DF achieves the strongest overall performance ($F_1\!=\!0.939$), while RF and Holmes surprisingly collapse with near zero $F_1$ scores. Figure~\ref{fig:ow-combined-au-ca-breakdown-grid} provides another view of this result by plotting $\pi_{10}$--recall curves across all three scenarios. The curves show that DF is the most consistent method: it sustains a strong precision--recall tradeoff in the baseline, in the cross-network setting, and under pooled training/testing, whereas RF and Holmes exhibit pronounced sensitivity under network mismatch.

This gap is notable because DF relies only on the direction sequence, whereas RF and Holmes depend more heavily on timing-sensitive representations (e.g., TAM-style aggregation and convolutional encoders), which are more susceptible to distribution shift induced by differing timing and bandwidth conditions across networks (we provide a detailed analysis on this issue in Section~\ref{exp:jitter}). Tik-Tok, which incorporates timing but retains DF’s direction-based sequence, degrades less severely than RF/Holmes. Finally, when we pool CA and AU traces in both training and testing, the degradation disappears and all methods recover strong performance, indicating that the cross-network failure is primarily driven by a train-test mismatch in network conditions. 

Overall, our results show that state-of-the-art WF attacks can achieve simultaneously high recall and $r$-precision at the guard against a real open world, supporting the conclusion that open-world WF remains a realistic threat.
It is worth noting that the highest $F_1$ score slightly increases by 0.01 compared to the baseline for pooled training, and slightly decreases by 0.025 for cross-network training.

\subsubsection{Time-Based Segmentation Without Guard Vantage}
\label{sec:no-guard-advantage}

To examine if attacking as a guard causes this increase, we skipped the part of our data sanitization steps that relies on circuit IDs, which are only visible to the guard. Instead, we relied on time-based segmentation for webpage traces (as detailed in Section~\ref{sec:time-based-seg}). 
The raw data remained the same; it was still collected at the guard, so it best represents an attacker on the wire between the client and the guard. 
Results are in Table~\ref{tab:ow-without-guard}. Focusing on the best-performing attack, DF, the $F_1$ score drops by 0.019 in the cross-network (train AU, test CA) scenario, and it drops by 0.005 in the pooled scenario. 
A comparison with the baseline shows that using the real Tor open world in our dataset for training and testing does not weaken classifiers compared to the standard laboratory setting with a synthetic open-world dataset.

\begin{table}[htbp]
\centering
\caption{Open-world performance without the guard advantage (no circuit-ID demultiplexing), reported in the same format as Table~\ref{tab:ow-with-guard} using $\pi_{10}$, $R$, and $F_1$ (thresholds maximize $F_1$).}
\label{tab:ow-without-guard}

\renewcommand{\arraystretch}{1.2}

\normalsize

\setlength{\tabcolsep}{4.5pt}

\begin{tabular}{l ccc ccc}
\toprule

\multirow{2}{*}{Classifier} 
    & \multicolumn{3}{c}{Train: AU / Test: CA} 
    & \multicolumn{3}{c}{Train \& Test: Both} \\
\cmidrule(lr){2-4} \cmidrule(l){5-7}

 & $\pi_{10}$ & $R$ & $F_1$ & $\pi_{10}$ & $R$ & $F_1$ \\
\midrule

$k$-FP    & 0.728 & 0.363 & 0.484 & 0.969 & 0.885 & 0.925 \\
DF        & \textbf{0.936} & \textbf{0.905} & \textbf{0.920} & \textbf{0.985} & \textbf{0.955} & \textbf{0.969} \\
Tik-Tok   & 0.923 & 0.856 & 0.888 & 0.977 & 0.942 & 0.959 \\
RF        & 0.003 & 0.000 & 0.001 & 0.973 & 0.960 & 0.966 \\
Holmes    & 0.161 & 0.002 & 0.004 & 0.955 & 0.950 & 0.952 \\
\bottomrule
\end{tabular}
\end{table}

\subsubsection{Impact of Open-World Scaling on Model Performance}
\label{sec:r_sensitivity}

\begin{figure*}[t]
    \centering
    \includegraphics[width=1\linewidth]{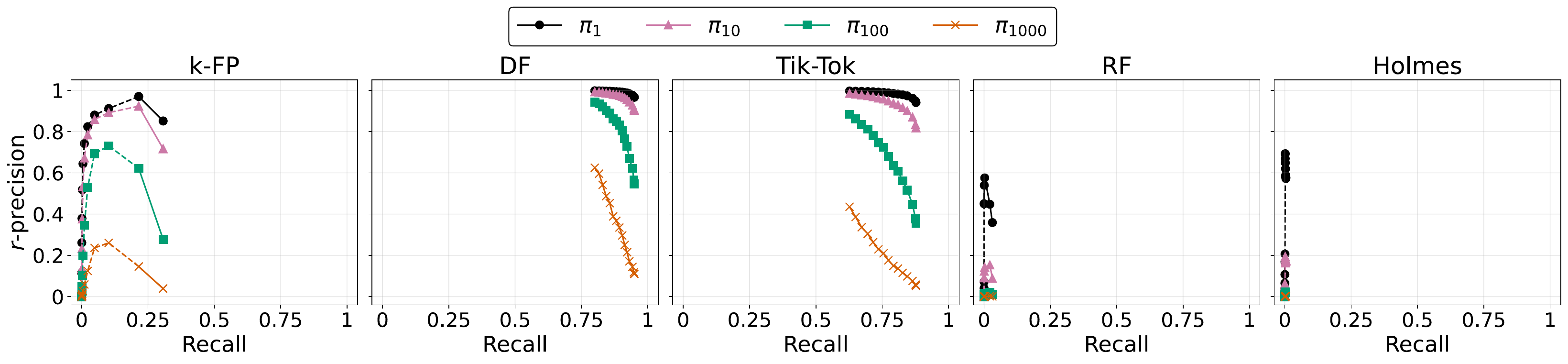}
    \caption{Performance comparison across different open-world ratios ($\pi_r$). As $r$ increases, the precision collapses, revealing the robustness of different architectures to unmonitored noise. Dashed lines (where applicable) represent $r$-precision corrections using the Wilson score interval when the number of false positives is below 10.}
    \label{fig:ow-combined-au-ca-rp-grid}
\end{figure*}

In this section, we provide a detailed analysis of how the scalability of WF models is affected by the ratio of unmonitored to monitored traffic ($r$). Figure~\ref{fig:ow-combined-au-ca-rp-grid} illustrates the $r$-precision (for different values of $r$) vs. recall curves for different models when trained on AU traces and tested on CA traces. We justify our selection of $\pi_{10}$ for model comparison as follows:

\paragraph{Benchmark Sensitivity.} 
As demonstrated in Figure~\ref{fig:ow-combined-au-ca-rp-grid}, in the balanced case ($\pi_{1}$), high-performing models may reach saturation (precision $\approx$1.0), making it difficult to distinguish between different configurations. By introducing a $10\times$ larger unmonitored world, we move the evaluation into a more discriminative regime where differences in the False-Positive Rate (FPR) significantly impact the precision score, thereby offering a more rigorous and reproducible metric for model comparison. Therefore, utilizing $\pi_{10}$ in our experiments ensures that our evaluation remains sensitive to architectural improvements.

\paragraph{Stability.}
As shown in the results for DF and Tik-Tok (Figure~\ref{fig:ow-combined-au-ca-rp-grid}), the performance curves for $\pi_{10}$ (pink triangles) closely follow the baseline $\pi_{1}$ curves. This suggests that $\pi_{10}$ serves as a stable benchmark that captures the model’s operational capacity before it encounters the rapid performance degradation observed at $\pi_{1000}$.

\subsubsection{Number of Monitored Training Traces}

\begin{figure}[htbp]
    \centering
    \includegraphics[width=1\linewidth]{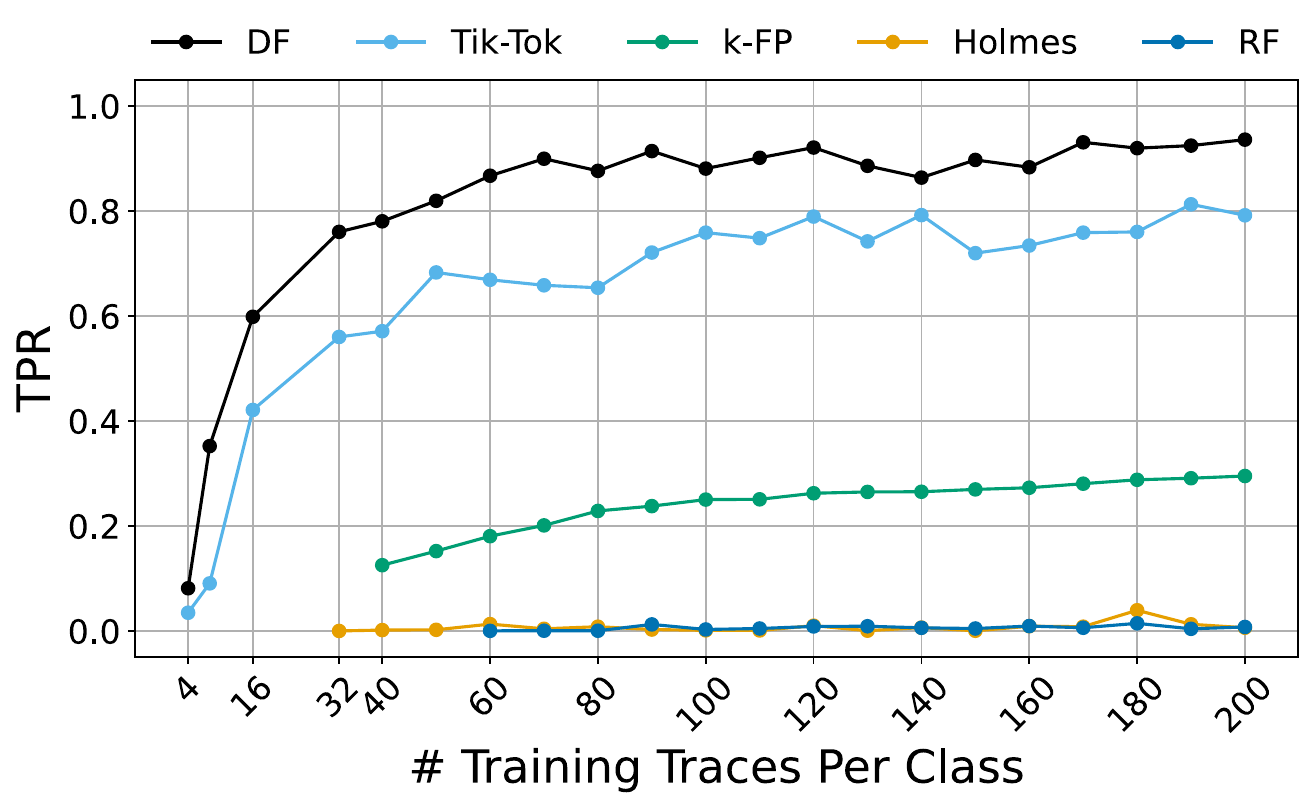}
    \caption{The effect of number of monitored training set size on TPR. We report TPR at a fixed target FPR of $0.5\%$ when training on AU monitored traces and testing on CA monitored traces. Missing points indicate that no threshold achieved FPR~$\le 0.5\%$ on the evaluation set.}
    \label{fig:ow-test-set-abl-mon-au_ca-tpr}
\end{figure}

We next examine how the monitored training set size impacts open-world performance under network mismatch. Figure~\ref{fig:ow-test-set-abl-mon-au_ca-tpr} reports the TPR at a fixed 0.5\% FPR as we vary the number of monitored traces per class (train AU, test CA). For each class, we sampled monitored training traces uniformly at random.
DF remains the leading attack, achieving 0.90 TPR with 70 traces per webpage; performance saturates thereafter (gaining only 0.04 at 200 traces). These results suggest that a guard-side adversary can achieve high-precision open-world detection with a relatively small number of training traces per monitored webpage.

\subsubsection{Per-Webpage Fingerprintability}

\begin{figure}[htbp]
    \centering
    \includegraphics[width=1\linewidth]{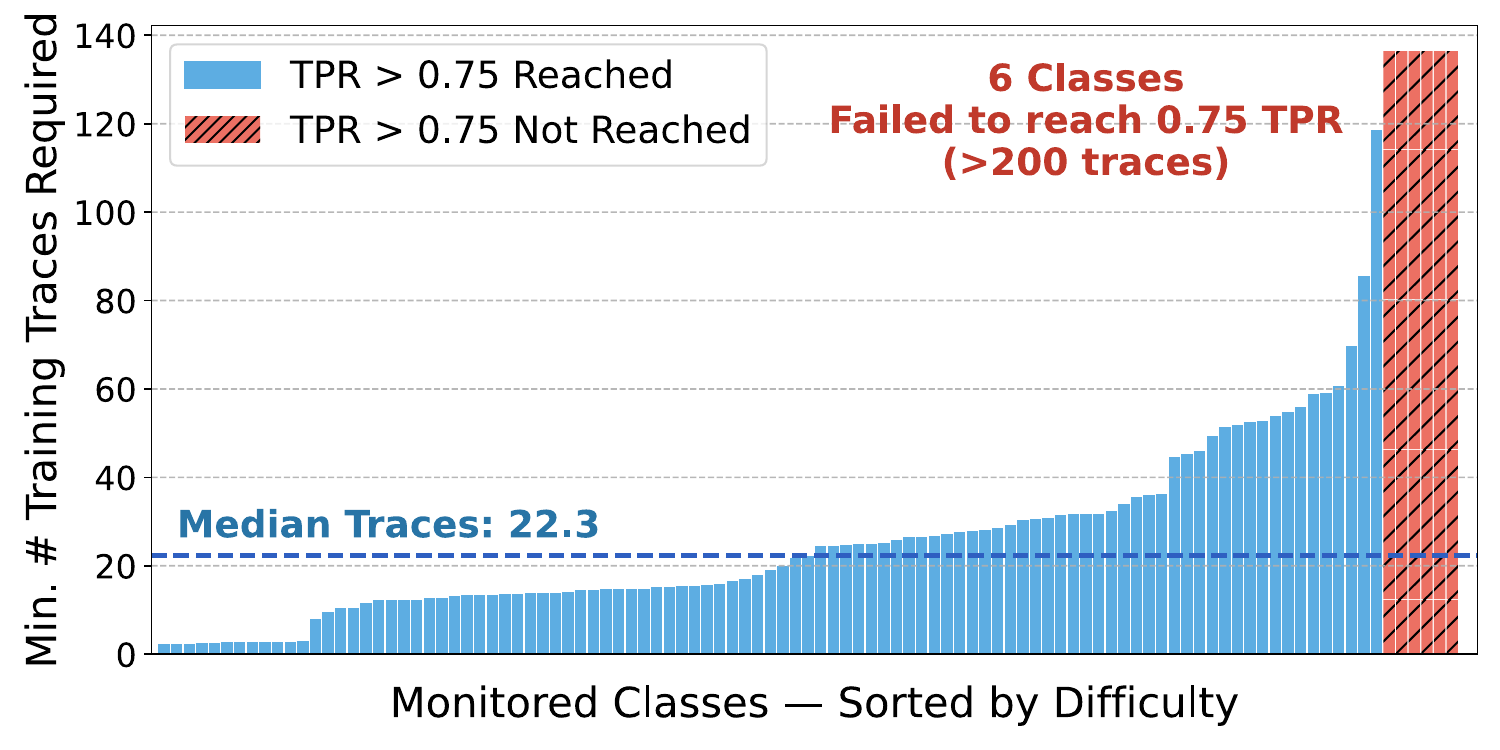}
    \caption{Per-webpage number of training instances for DF to achieve $\mathrm{TPR}>0.75$ under cross-network evaluation (train AU, test CA). Bars show, for each monitored class, the minimum number of labeled monitored training traces needed to reach the target, capped at 200; hatched bars indicate classes that do not reach $\mathrm{TPR}>0.75$ within this budget.}
    \label{fig:ow-test-set-abl-mon-au_ca-tpr-cm}
\end{figure}

Aggregate TPR often conceals variability in classification performance across monitored webpages. Therefore, we next examine the minimum number of traces required per webpage to reach a target TPR. Focusing on DF, we varied the per-class training budget (AU traces) from 2 to 200 and evaluated on the full set of CA traces.
Figure~\ref{fig:ow-test-set-abl-mon-au_ca-tpr-cm} reports the minimum number of training traces required for each monitored class to achieve $\mathrm{TPR}>0.75$ (sorted by difficulty). 
About 48\% of our monitored classes satisfy this threshold with fewer than 20 training traces. With a budget of 60 traces, over 91\% of monitored classes achieve the target TPR. Notably, six of our monitored webpages cannot reach the TPR of 0.75 even with 200 training traces. 
These webpages include region-specific adult content, a regional gaming platform, a non-governmental organization, and a technical reference webpage. Due to the nature of these webpages, it is possible that they were inadvertently included in the non-monitored training set, leading to an elevated FNR.

\subsubsection{Per-Webpage Error Rate Analysis}
\label{sec:per-webpage-analysis-fnr}

To better understand the fingerprintability of individual webpages, we analyze the distribution of error rates across monitored webpages. In the open-world setting, attackers typically prioritize a low false-positive rate by raising the decision threshold that distinguishes monitored from non-monitored traffic. However, this conservatism comes at a cost: as the threshold increases, more monitored traces are rejected as non-monitored, thereby increasing the false-negative rate (FNR). We therefore study how this precision--recall trade-off manifests across monitored webpages by examining the per-page error distribution under fixed operating points.

\begin{figure}[t!]
    \centering
    \includegraphics[width=1\linewidth]{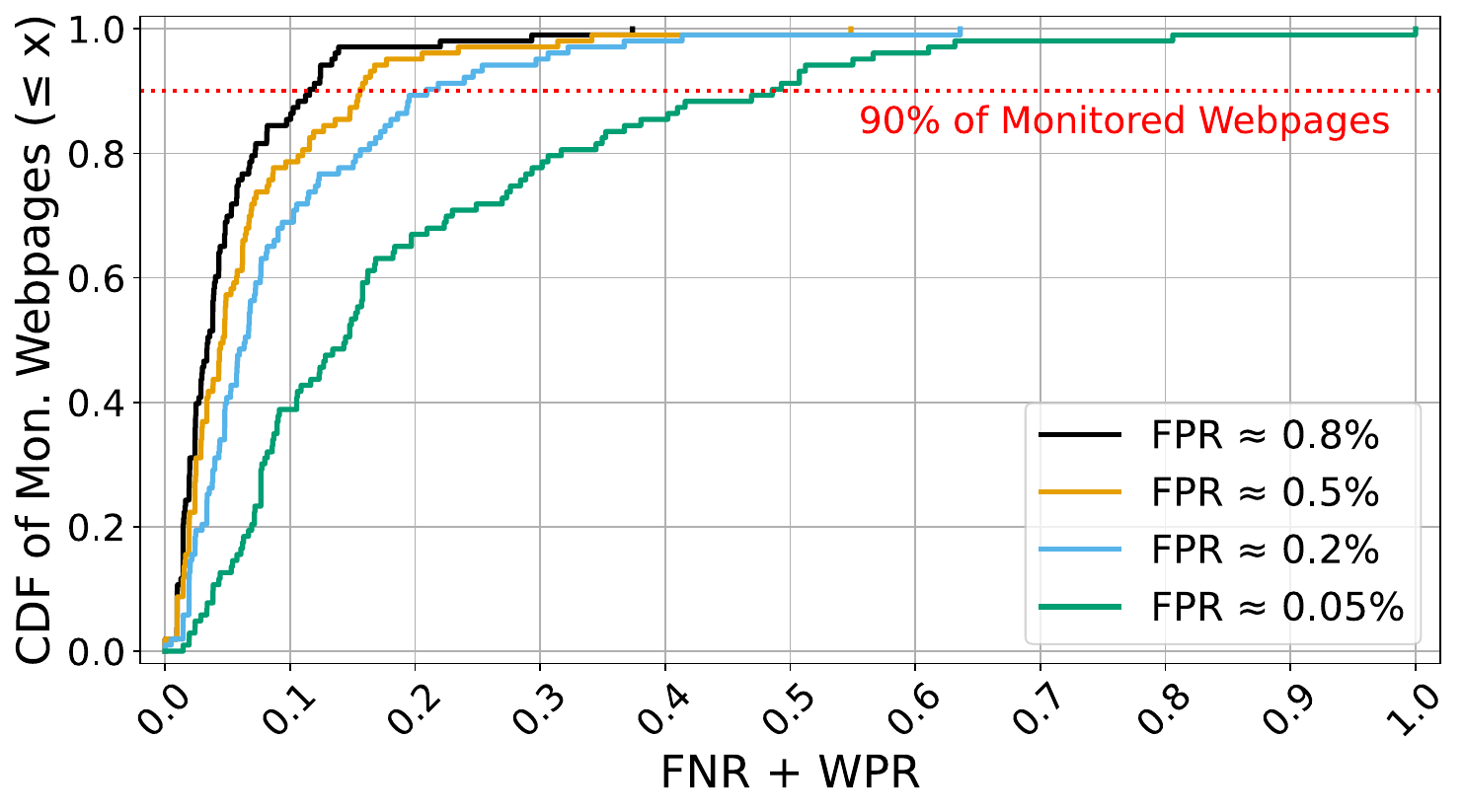}
    \caption{Cumulative Distribution of False Negative Rate (FNR) and Wrong Positive Rate (WPR) for DF. The model was trained on traces from the AU client and evaluated on traces from the CA. Each curve represents a different classification threshold, tuned for a specific target False-Positive Rate (FPR). The shift toward the right as FPR decreases illustrates the inherent trade-off between precision and leakage.} \label{fig:ow-test-set-ca_au-cm-df-fnr-ecdf}
\end{figure}

Figure~\ref{fig:ow-test-set-ca_au-cm-df-fnr-ecdf} plots the empirical CDF of the combined per-webpage error rate (FNR + WPR) for the DF classifier when trained on traces from the AU client and tested on CA traces. Two trends stand out. First, for moderate FPR targets, most monitored webpages remain highly fingerprintable. At a target FPR of $\approx$0.5\% (orange curve), more than 78\% of monitored webpages have an error rate below 10\%, and over 97\% remain below 25\%. Second, tightening the FPR shifts the distribution rightward, quantifying the expected open-world trade-off. Yet, even at our most conservative operating point (FPR $\approx$0.05\%), more than 71\% of monitored webpages still achieve error rates below 25\%. This indicates that leakage is concentrated among a substantial subset of pages.

Although DF attains FPR $\approx$0.05\% in this specific evaluation, not all classifiers and experimental regimes considered in this paper can reliably operate at such a stringent rate. We therefore adopt FPR $=0.5\%$ as our baseline operating point for cross-model comparisons.

\subsubsection{Number of Non-Monitored Training Traces}

A practical open-world WF classifier should benefit from additional non-monitored training data, since richer background traffic helps the model suppress false alarms at a fixed detection level. We measured how DF’s false-positive rate evolves as we increased the number of non-monitored training traces while holding the monitored training set fixed.

\begin{figure}[t!]
    \centering
    \includegraphics[width=1\linewidth]{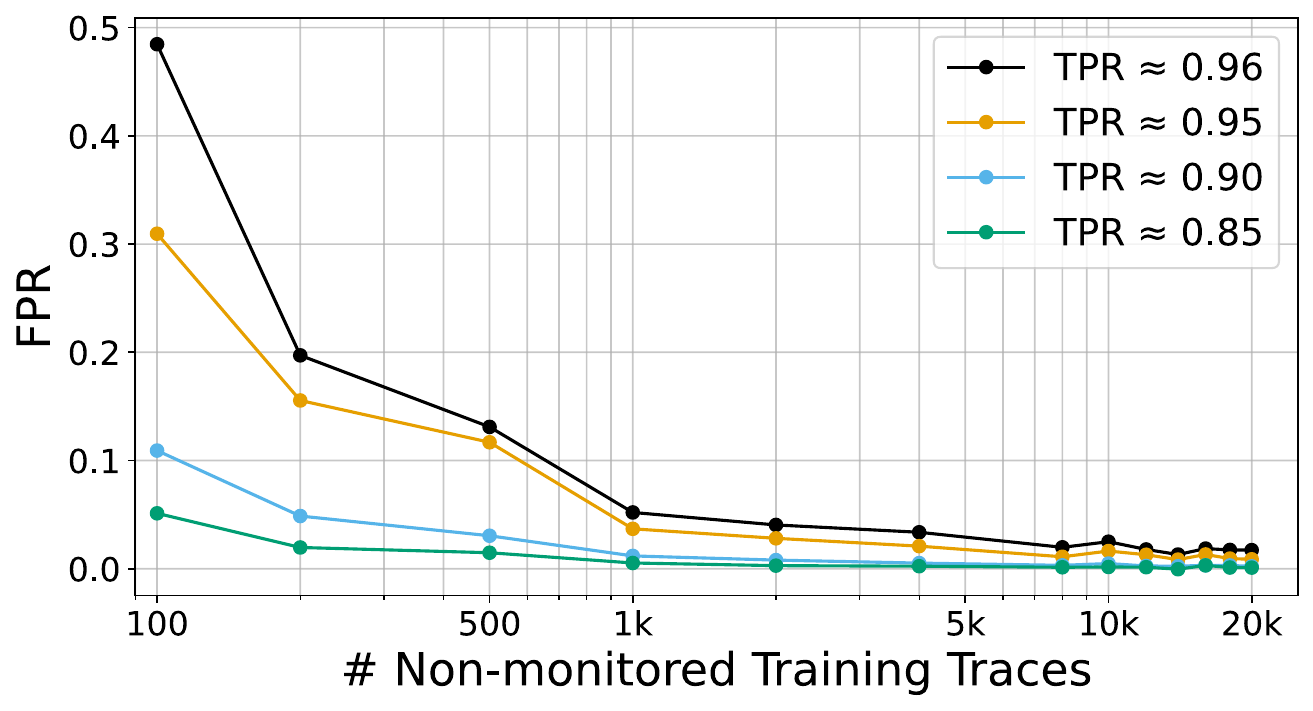}
     \caption{Effect of non-monitored training set size on open-world false positives for DF (train AU, test CA). We fix the monitored training set to 200 traces per class and vary the number of non-monitored training traces; the target TPRs obtained by threshold sweeping on the evaluation set.}
    \label{fig:ow-test-set-abl-unmon-au_ca-fpr}
\end{figure}

We trained DF on AU traces (200 monitored traces per class) and varied the size of the non-monitored training set; we then evaluated on CA monitored traces together with a subsequent non-monitored test set.
Figure~\ref{fig:ow-test-set-abl-unmon-au_ca-fpr} reports the achieved FPR at several target true-positive rates (TPR), obtained by sweeping the decision threshold. FPR drops sharply up to $\approx$1,000 non-monitored traces, followed by diminishing but valuable gains. At a fixed TPR of 0.90, expanding the non-monitored training set from 2,000 to 20,000 traces reduces FPR from $\approx$0.008 to $\approx$0.002. This reduction is critical for low-base-rate scenarios, particularly for larger $r$, where $\pi_{100}$ improves from $\approx$0.530 to $\approx$0.802.

\subsubsection{Sensitivity to Network Timing}
\label{exp:jitter}

Table~\ref{tab:ow-with-guard} shows that while several attacks perform strongly on pooled data, some methods degrade substantially under cross-network evaluation (e.g.\ train AU, test CA) due to their reliance on timing. 
To verify this, we conducted a controlled \emph{jitter} experiment.
First, we pooled the monitored dataset (CA+AU). 
While training traces remain unmodified, we injected additional delay into the \emph{monitored} test traces. 
Specifically, we added an i.i.d.\ jitter to the inter-arrival times (IATs) of the cells, sampled uniformly from $[0,J]$~ms. 
Because this jitter is cumulative, it increases the overall trace duration; for instance, with $J=20$~ms and a trace of 5{,}000 cells, the expected extension is roughly $5{,}000\times 10$~ms $\approx 50$~s. 
To maintain realistic conditions, we truncated test traces that exceeded the maximum duration of 45~s. 
All models were evaluated at FPR $=0.5\%$.

\begin{figure}[htbp]
    \centering
    \includegraphics[width=1\linewidth]{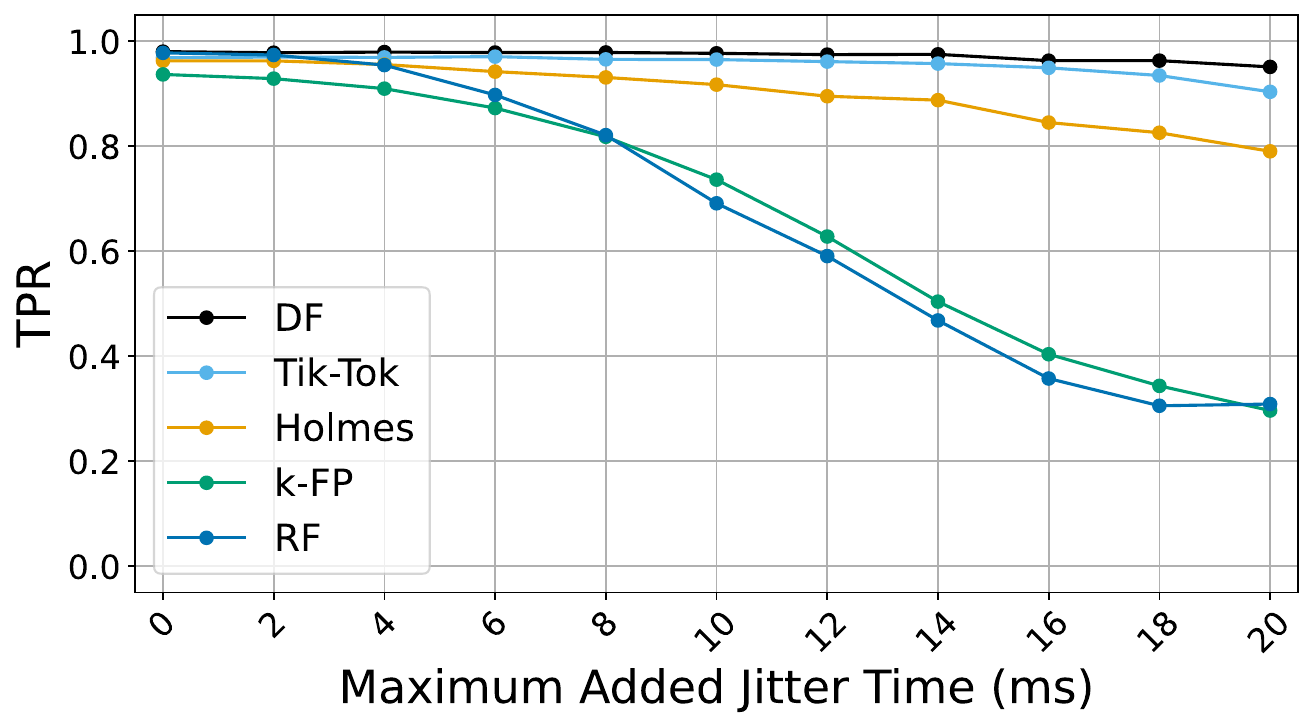}
    \caption{Sensitivity of WF attacks to timing jitter. Models are trained on unmodified CA+AU data. For testing, we inject i.i.d.\ jitter into the \emph{monitored} traces by adding a random delay uniformly sampled from $[0, J]$ to each cell’s inter-arrival time.}
    \label{fig:ow-combined-jitter_low-fpr-tpr}
\end{figure}

Figure~\ref{fig:ow-combined-jitter_low-fpr-tpr} reports the resulting TPR as a function of the maximum jitter $J$. The attacks exhibit markedly different robustness profiles. At $J=20$~ms, DF only loses $\approx$0.03 TPR, as it relies only on direction sequences and the loss is due to truncation. 
Tik-Tok also shows stability across the full range of jitter values (TPR loss of $\approx$0.06).
In contrast, timing-heavy methods degrade steadily: Holmes shows a gradual decline (from $\approx$0.96 to $\approx$0.79 at $J=20$~ms), while $k$-FP drops more sharply to $\approx$0.30. RF appears to be highly sensitive as well, with TPR falling from $\approx$0.98 to $\approx$0.31 at $J=20$~ms. 
Their sensitivity to timing jitter can explain why they did not perform well on cross-network training, as the guard's latency to CA and AU differs significantly. 
We further explore the effect of timing and latency on RF's Traffic Aggregation Matrix (TAM) representation in Appendix~\ref{app:rf-hyper}.

\subsubsection{Early-Stage Fingerprinting}
\label{sec:early-detection}

\begin{figure}[htbp]
    \centering
    \includegraphics[width=1\linewidth]{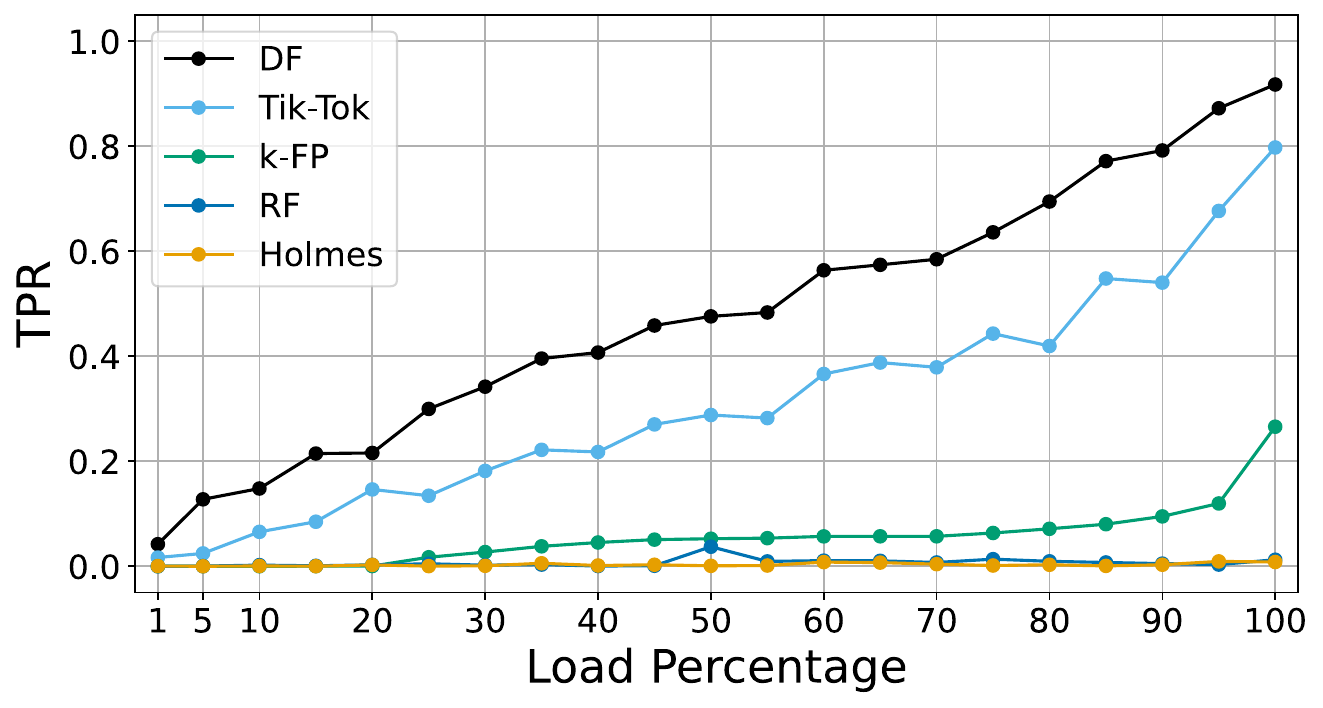}
    \caption{Early-stage open-world performance under network mismatch (train AU, test CA). We truncate each monitored test trace to the first $x\%$ of its cells (``load percentage'') and report the resulting TPR at a fixed target FPR of $0.5\%$.}
    \label{fig:ow-test-set-load-percent_low-fpr-tpr}
\end{figure}

Early-stage fingerprinting allows an adversary to act on a prediction before a full page load completes (e.g., to support timely monitoring or blocking). Motivated by recent early-stage WF designs~\cite{deng2024robust}, we evaluated whether WF attacks could make reliable cross-network predictions with only the initial portion of each trace. As such, under cross-network evaluation, we trained the models on the full AU dataset and then performed early-detection evaluation on the CA dataset by truncating each \emph{monitored} test trace to the first $x\%$ of its length (the \emph{load percentage}). Figure~\ref{fig:ow-test-set-load-percent_low-fpr-tpr} reports the TPR as $x$ increases from 1\% to 100\% (FPR is fixed at 0.5\%).

\begin{table*}[t!]
    \centering
    \caption{The effect of concept drift on different models. The models are trained on the baseline (month 0) and evaluated on month 2 and month 6 (same vantage point). We report $r$-precision at $r{=}10$ ($\pi_{10}$), recall ($R$), and the $F_1$ score. The decision threshold is tuned to maximize the $F_1$ score on month 0.}
    \label{tab:concept-drift}
    
        \normalsize
        \renewcommand{\arraystretch}{1.2}
    
    \begin{tabular}{@{} l ccc ccc ccc @{}}
    \toprule
    \multirow{2}{*}{Classifier} 
        & \multicolumn{3}{c}{Month 0} 
        & \multicolumn{3}{c}{Month 2}
        & \multicolumn{3}{c}{Month 6} \\ 
    \cmidrule(lr){2-4} \cmidrule(lr){5-7} \cmidrule(lr){8-10} 
    
     & $\pi_{10}$ & $R$ & $F_1$ 
     & $\pi_{10}$ & $R$ & $F_1$ 
     & $\pi_{10}$ & $R$ & $F_1$ \\ 
    \midrule
    $k$-FP    & 0.946 & 0.908 & 0.926 & 0.758 & 0.764 & 0.761 & 0.632 & 0.482 & 0.547 \\
    DF        & 0.985 & 0.950 & 0.967 & 0.910 & 0.805 & 0.854 & 0.836 & 0.580 & 0.685 \\
    Tik-Tok   & 0.972 & 0.941 & 0.956 & 0.840 & 0.814 & 0.827 & 0.812 & 0.546 & 0.653 \\
    RF        & \textbf{0.989} & \textbf{0.955} & \textbf{0.972} & \textbf{0.941} & \textbf{0.875} & \textbf{0.907} & \textbf{0.857} & \textbf{0.673} & \textbf{0.754} \\
    Holmes    & 0.974 & 0.957 & 0.966 & 0.796 & 0.692 & 0.740 & 0.702 & 0.558 & 0.622 \\
    
    \bottomrule
    \end{tabular}
\end{table*}

The attacks perform poorly in this cross-network setting. 
The strongest attack, DF, requires 90\% of data to achieve only $\approx$0.8 TPR.
Notably, Holmes, which is explicitly designed for early detection, yields a near-zero TPR throughout. 
This suggests that timing- and distribution-sensitive representations do not transfer well when network conditions shift and would require more thorough training than a single vantage point. 

\subsubsection{Robustness to Concept Drift}
\label{sec:concept-drift}

Prior work has shown that website fingerprinting models degrade over time as webpages evolve, a phenomenon commonly referred to as \emph{concept drift}~\cite{wang2016realistically,wang2022snwf}. To measure the resilience of our methodology to this drift under realistic deployment conditions, we conducted a longitudinal study using our UK client. We collected independent datasets (both monitored and non-monitored) at three time points: an initial baseline (month 0), two months later, and six months later. We trained each model solely on the baseline data and evaluated it on the subsequent datasets from the same vantage point.

Table~\ref{tab:concept-drift} summarizes the results for different classifiers. As expected, performance degrades due to drift for all attacks. In particular, RF shows strong initial stability: starting with an $F_1$ score of 0.972, it maintains 0.907 after two months, before decreasing to 0.754 after six months. The decline is driven mainly by recall, indicating that an increasing fraction of monitored traces are missed as pages evolve. In contrast, precision degrades more moderately ($\pi_{10}$ decreases by 0.132 and FPR rises by only $\approx$0.008 from near zero), suggesting that positive predictions remain mostly reliable. Notably, although RF is less robust under cross-network shift (Table~\ref{tab:ow-with-guard}), it outperforms other classifiers in this experiment, suggesting that robustness to network mismatch and robustness to temporal evolution are distinct properties. 

\subsection{Website Fingerprinting Under Conflux}
\label{sec:conflux-exp}

We next evaluate website fingerprinting under Conflux using our post-Conflux dataset (Table~\ref{tab:overall-traces}). Recall that under Conflux, a webpage load may be split across two linked circuits (legs); in this experiment we model a guard-side adversary that observes only a single leg of each Conflux connection. 
Using the same open-world background construction as in Section~\ref{sec:exp:ow-preconflux}, we analyze Conflux traces. 
We trained on Conflux traces from the AU or UK client and tested on those from the CA client. 
We report $\pi_{10}$, recall ($R$), and the corresponding $F_1$ score (computed from $\pi_{10}$ and $R$), selecting the decision threshold that maximizes $F_1$.

Table~\ref{tab:cfx-single-col} shows a marked degradation compared to the pre-Conflux cross-network results in Table~\ref{tab:ow-with-guard}. 
$F_1$ score of DF drops from 0.939 previously to 0.379 under Conflux, and that of Tik-Tok similarly decreases from 0.872 to 0.297. 
This is due to Conflux fragmenting the guard’s view of a page load.
Our client-side ground-truth traces indicate that primary-leg switches can occur repeatedly (often after transmitting only $\approx$50~KB).\footnote{This number is related to the default Conflux scheduling algorithm, LowRTT, and the frequency of SENDME cells (every 100 cells $\approx$50~KB).}
In our measurements, this effect is pronounced: approximately 65\% of monitored traces contain less than half of the full page-load cells, indicating that our guard was frequently not the favored leg under Conflux scheduling. 

Moreover, while most methods perform substantially worse in the Conflux setting compared to the non-Conflux baseline, RF demonstrates comparatively stronger performance when trained on the UK dataset and tested on the CA dataset (Table~\ref{tab:cfx-single-col}). 
This suggests that the TAM-based representation remains robust when the training and testing data share similar network conditions. 
Specifically, the latency difference between the UK and CA clients relative to the guard node is $\approx$10~ms, much less than the $\approx$150~ms difference between AU and CA clients. Overall, these results indicate that, although Conflux was not designed as a WF defense, it introduces significant challenges for guard-side attackers capable of observing only one leg of a Conflux connection.

\begin{table}[t!]
\centering
\setlength{\tabcolsep}{4.5pt}
\renewcommand{\arraystretch}{1.2}

\caption{Open-world WF performance under Conflux (single-leg guard observation). We report $r$-precision at $r{=}10$ ($\pi_{10}$), recall ($R$), and $F_1$ score ($F_1$) calculated using $\pi_{10}$. For each model, the decision threshold is selected to maximize $F_1$.}

\label{tab:cfx-single-col}

\begin{tabular}{l ccc ccc} 
\toprule

\multirow{2}{*}{Classifier} 
 & \multicolumn{3}{c}{Train: AU / Test: CA} 
 & \multicolumn{3}{c}{Train: UK / Test: CA} \\
\cmidrule(lr){2-4} \cmidrule(lr){5-7}

 & $\pi_{10}$ & $R$ & $F_1$ & $\pi_{10}$ & R & $F_1$ \\
\midrule

$k$-FP    & 0.130 & 0.090 & 0.107 & 0.387 & 0.228 & 0.287 \\
DF      & \textbf{0.558} & \textbf{0.287} & \textbf{0.379} & 0.616 & 0.285 & 0.389 \\
Tik-Tok & 0.399 & 0.237 & 0.297 & 0.563 & 0.369 & 0.446 \\
RF      & 0.009 & 0.002 & 0.003 & \textbf{0.537} & \textbf{0.557} & \textbf{0.547} \\
Holmes  & 0.018 & 0.011 & 0.013 & 0.378 & 0.346 & 0.361 \\
\bottomrule
\end{tabular}%
\end{table}

\subsubsection{Conflux Scheduling Algorithm}
\label{sec:cfx-alg}

In Tor’s default configuration, Conflux uses \textsc{LowRTT} scheduling: endpoints prefer sending cells on the leg with the lowest measured RTT, provided the connection is in the unblocked state and the leg has available congestion-window capacity~\cite{tor-prop329}. 
Because RTT reflects both network path characteristics and relay-side queuing, this design naturally biases data transfer toward the leg that offers lower delay. 
Consequently, we hypothesize that a ``powerful'' guard (i.e., a guard leg with consistently lower RTT) is more likely to be selected as the primary leg during the initial phase of a connection. 
Critically, such a guard would therefore be more likely to observe the beginning of the trace rather than only a later, schedule-dependent fragment. 
Furthermore, such a guard can see a larger fraction of the total page load on its leg.

\subsubsection{Conflux First-Segment Detection}
\label{sec:fs-detection}

The initial portion of the trace is especially valuable for website fingerprinting as it is feature-rich~\cite{gong2020zero,deng2024robust}. Therefore, we define a \emph{first-segment} (FS) trace as a Conflux leg trace where the leg is selected as the \textit{primary leg} by both the client \textit{and} the exit relay during the initial loading of a webpage. To test whether first-segment visibility explains the degradation we observe under single-leg Conflux observation, we first designed a lightweight guard-side detector that classifies traces as first-segment (FS) or non-FS:

\paragraph{Guard-Side FS Detector.}
Based on natural traffic patterns observed during webpage loading, we constructed a guard-side FS detector that enables an attacker controlling a guard node to identify FS traces. The detector classifies a trace as an FS trace if it satisfies three conditions: (1) the first cell following the Conflux handshake is outgoing; (2) at least one incoming cell appears within the first 10 cells; and (3) at least one outgoing cell appears within a 10-cell window immediately following the first incoming cell. The window sizes for this detector were tuned empirically using traces gathered from our clients.

\begin{algorithm}
\caption{Ground Truth Primary Leg Identification from Client Traces}
\label{alg:client_ground_truth}
\begin{algorithmic}[1]
\Require Conflux Set with two legs: $\mathit{Leg}_A, \mathit{Leg}_B$,
\Statex where each Leg is a sequence of cells $\{c_1, c_2, \dots\}$.
\Statex Each cell $c$ has properties $\mathit{type}(c)$ and $\mathit{dir}(c)$.
\Ensure The primary leg of the Client and the Exit at the start of data transfer.

\Statex
\Statex \textbf{Step 1: Remove Conflux Handshake}
\Statex \Comment{Handshake ends at \texttt{CFX\_LINKED\_ACK} (21)}
\State $\mathit{Leg}_A \gets \text{StripHandshake}(\mathit{Leg}_A)$
\State $\mathit{Leg}_B \gets \text{StripHandshake}(\mathit{Leg}_B)$

\Statex
\Statex \textbf{Step 2: Identify Client Primary Leg}
\Statex \Comment{Identify leg containing \texttt{RELAY\_BEGIN} (1)}
\If{$\mathit{type}(\mathit{Leg}_A[0]) = 1$}
    \State $\mathit{ClientPrimary} \gets \mathit{Leg}_A$
    \State $\mathit{ClientSecondary} \gets \mathit{Leg}_B$
\ElsIf{$\mathit{type}(\mathit{Leg}_B[0]) = 1$}
    \State $\mathit{ClientPrimary} \gets \mathit{Leg}_B$
    \State $\mathit{ClientSecondary} \gets \mathit{Leg}_A$
\Else
    \State \Return \textbf{Unused Set}
\EndIf

\Statex
\Statex \textbf{Step 3: Identify Exit Relay Primary Leg}
\Statex $\triangleright$ \textit{Note: The client cannot send \texttt{RELAY\_DATA} (2) unless}
\Statex \textit{it has already received \texttt{RELAY\_CONNECTED} (4) from the exit relay.}
\Statex \Comment{Skip \texttt{RELAY\_BEGIN}s}
\State Let $c_{target}$ be the first cell in $\mathit{ClientPrimary}$ such that $\mathit{type}(c_{target}) \neq 1$.

\If{$\mathit{dir}(c_{target}) = \text{+1} \textbf{ and } \mathit{type}(c_{target}) = 2$}
    \State $\mathit{ExitPrimary} \gets \mathit{ClientSecondary}$
\Else
    \State $\mathit{ExitPrimary} \gets \mathit{ClientPrimary}$
\EndIf

\Statex
\State \Return $(\mathit{ClientPrimary}, \mathit{ExitPrimary})$
\end{algorithmic}
\end{algorithm}

\paragraph{Validation via Ground Truth.}
To evaluate the detector's accuracy, we derived ground truth data from our modified Tor clients. For this, we relied solely on cell types and directionality, eliminating the need to decrypt cell payloads. By inspecting specific control cells (such as \texttt{RELAY\_BEGIN} (Type 1) and \texttt{RELAY\_DATA} (Type 2)) we can definitively determine whether a specific trace qualifies as a First-Segment (FS) trace. Algorithm~\ref{alg:client_ground_truth} details the ground truth identification procedure. This algorithm proceeds in three steps:

\begin{enumerate}

\item We isolate the data transfer phase. As defined in Tor Proposal 329~\cite{tor-prop329}, the Conflux handshake concludes when the client sends a \texttt{CFX\_LINKED\_ACK} (Type 21) cell.\footnote{The exact technical names of these constants are shortened for clarity.} Consequently, we strip the handshake by discarding all cells up to and including the first occurrence of this cell on each leg.

\item We identify the client's primary leg. Following the handshake, the client initiates the stream by sending a \texttt{RELAY\_BEGIN} (Type 1) cell. We designate the leg carrying this initial command as the \textit{Client Primary Leg}. Note that multiple \texttt{RELAY\_BEGIN} cells may occur (e.g., due to timeouts).

\item We deduce the exit relay's primary leg. According to the Tor Specification~\cite{tor-spec}, a client must receive a \texttt{RELAY\_CONNECTED} (Type 4) cell from the exit before it can transmit \texttt{RELAY\_DATA} (Type 2). We utilize this dependency to infer the exit's leg choice. We examine the first cell on the \textit{Client Primary Leg} that is not a \texttt{RELAY\_BEGIN} cell. If this cell is an outgoing \texttt{RELAY\_DATA} cell, it implies the client received the required \texttt{RELAY\_CONNECTED} signal via the \textit{secondary} leg. In this case, we identify the secondary leg as the exit's primary leg. Conversely, if the sequence contains an exit relay response such as \texttt{RELAY\_CONNECTED} (or other control cells such as \texttt{CFX\_SWITCH}), the exit's choice aligns with the client's. This method proved robust in our testing, correctly handling edge cases such as early leg switching by the exit relay.

\end{enumerate}

\paragraph{Results.}
We compared our guard-side FS detector against the client-side ground truth (Algorithm~\ref{alg:client_ground_truth}) by matching client traces to guard traces via circuit IDs. We determined that the proposed detector achieves an accuracy of 99.7\% in correctly identifying FS traces from the guard side.

\subsubsection{Conflux Under a Powerful Guard}

We next determine whether a powerful guard is more likely to succeed in WF.
We simulated a latency advantage mentioned in Section~\ref{sec:cfx-alg} by manipulating RTT to \emph{other} guards using the Linux network emulator, \texttt{netem}. 
Specifically, on our CA client,
we added symmetric delay (to both inbound and outbound traffic) for all candidate guard IP addresses \emph{except} our own guard. We repeated Conflux data collection and open-world evaluation under these RTT conditions to quantify how increased trace coverage translates into attack performance.

\begin{figure}[htbp]
    \centering
    \includegraphics[width=1\linewidth]{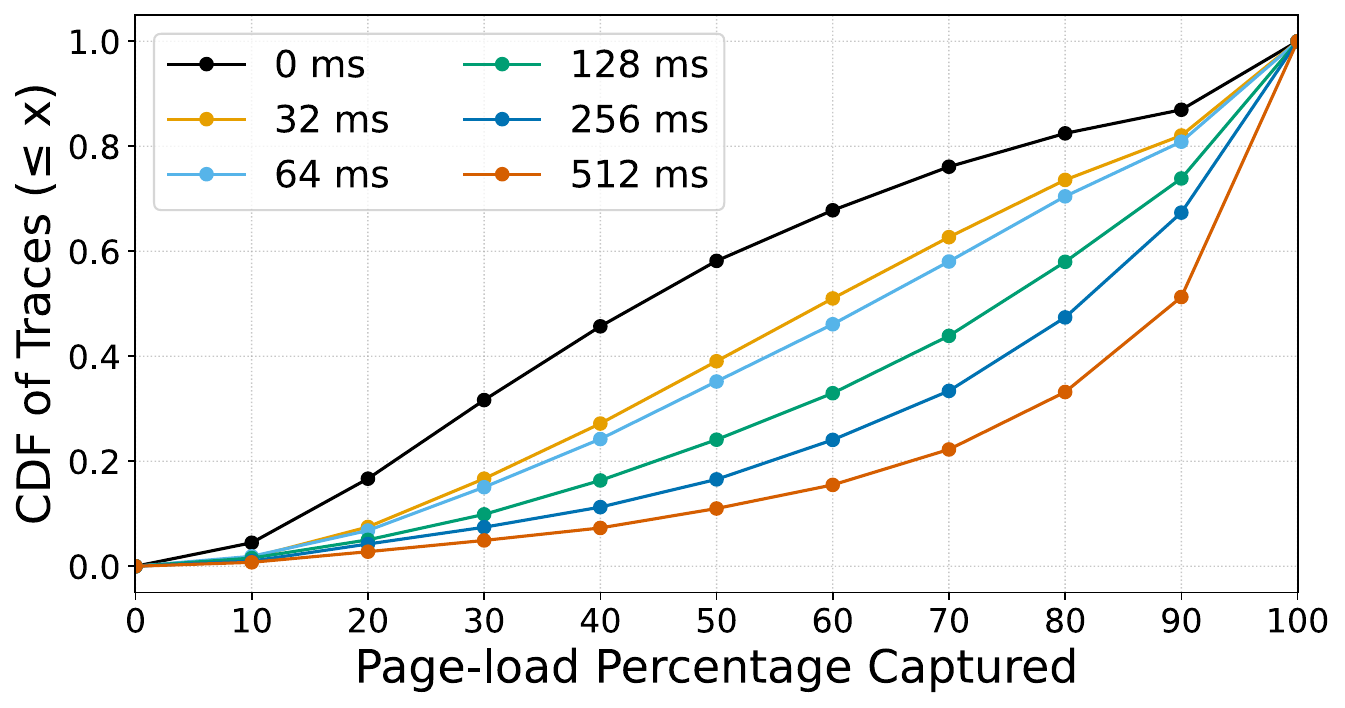}
    \caption{Trace coverage under Conflux vs.\ added RTT (CA client): CDF of the fraction of each full page-load trace observed at our guard.}
    \label{fig:cfx-figure-1}
\end{figure}

Figure~\ref{fig:cfx-figure-1} quantifies how added RTT to competing guards affects the fraction of each full page load captured by our guard (measured relative to the complete trace observed at the client). As the added RTT increases, the distribution shifts to the right: our guard captures a larger portion of each webpage load. For example, when no added delay is present, only $\approx$17\% of our CA monitored traces achieve the 80\% or higher coverage threshold. However, this number increases to $\approx$67\% when 512~ms is added to competing guards. Importantly, even a moderate and practically achievable latency advantage has a pronounced effect: at 128~ms added RTT, our guard captures a median of approximately 76\% of each webpage load (compared to $\approx$42\% in no added delay condition).

\begin{figure}[t]
    \centering
    \includegraphics[width=1\linewidth]{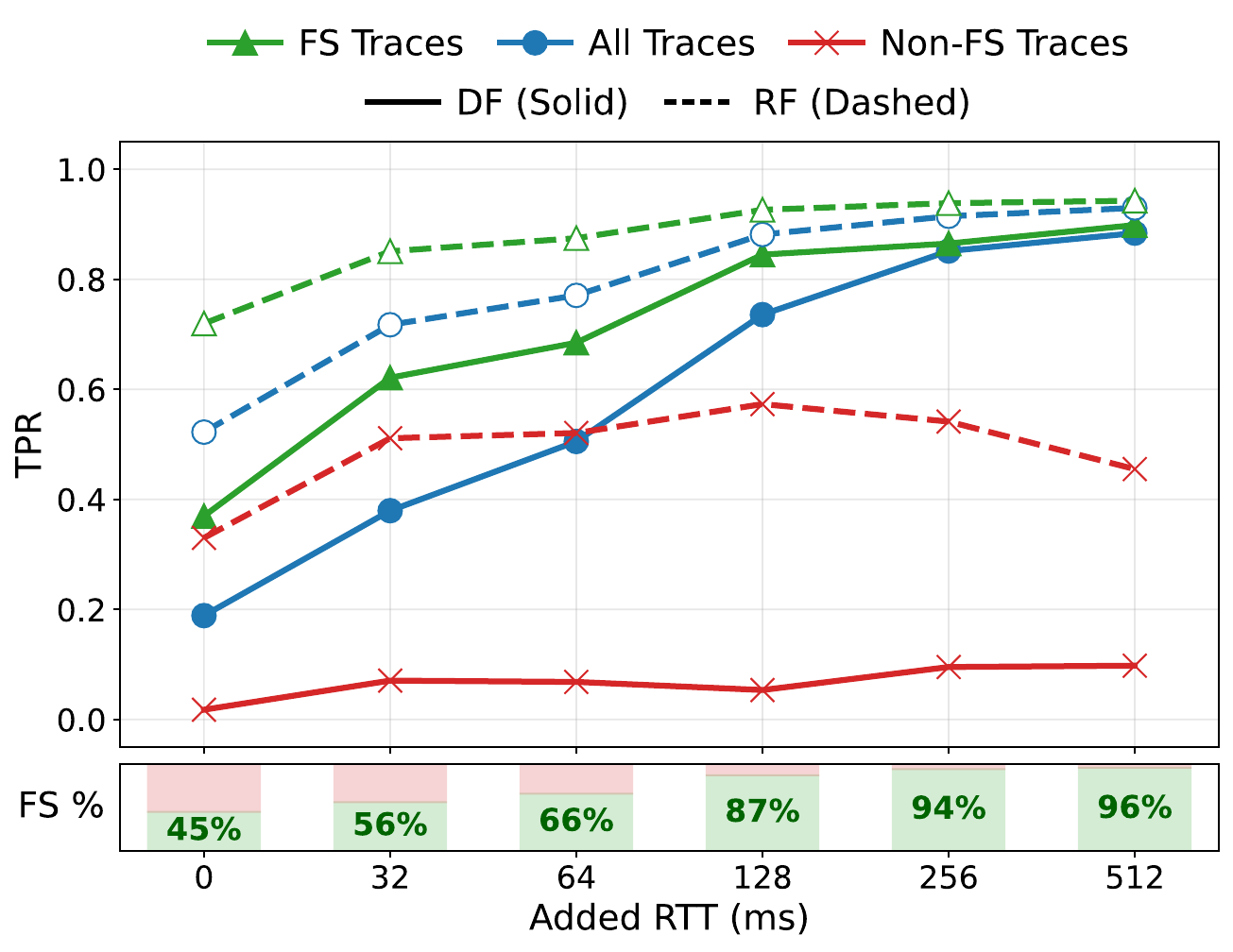}
    \caption{Conflux ``powerful guard'' simulation (CA client): open-world TPR of DF (solid) and RF (dashed) at FPR $=0.5\%$ versus added RTT to competing guards, shown for all traces and split into FS vs.\ non-FS subsets; the bottom bar reports the FS fraction at each added RTT.}
    \label{fig:cfx-ca-latency_cfx-combined-fs-grid}
\end{figure}

Next, we investigate the distribution of FS traces under these conditions and evaluate their specific contribution to classification performance. Figure~\ref{fig:cfx-ca-latency_cfx-combined-fs-grid} shows the results of this experiment for the DF and RF classifiers. 
This figure summarizes both (i) the fraction of traces that our guard observes as first-segment (FS) and (ii) the resulting TPR at FPR $=0.5\%$, also breaking down TPR for FS vs. non-FS traces.

The FS fraction, reported beneath the plot, increases monotonically with added RTT, from $\approx$45\% at 0~ms to $\approx$96\% at 512~ms, confirming that \textsc{LowRTT} increasingly selects our guard leg as primary during initial transfer as competing legs become slower. The results show that increased WF performance is due almost exclusively to the increased proportion and classification ease of FS traces. In contrast, TPR on non-FS traces changed little with added RTT. For DF, overall TPR increases from 0.189 at 0~ms (no latency advantage) to 0.736 at 128~ms (a moderate and practically achievable latency advantage). This result significantly exceeds the non-FS TPR increase from 0.018 to 0.053 in the same range.

These results suggest that a guard adversary with a significant RTT advantage can recover much of the fingerprinting power lost under single-leg Conflux observation. 
Such an advantage is plausible in practice given Tor’s small, long-lived guard set: if an adversarial guard is topologically closer to the client than the client’s other guards, \textsc{LowRTT} scheduling will consistently route early traffic over the adversarial leg. 

Finally, we note that while all models benefit most from FS traces, RF is comparatively less impaired on non-FS traces than DF (maintaining $\approx$0.330--0.573 TPR compared to DF's $<0.1$ on that subset), suggesting it can extract more usable signal from partial, schedule-fragmented Conflux legs and achieve a higher TPR on these traces. As a result, RF achieves the overall TPR of $\approx$0.881 at 128~ms, an increase of 0.145 compared to DF for the same latency.

\section{Discussion}

\label{sec:discussion}
\subsection{Why Did WF Succeed in Our Setting?}

The WF attacks successful in our setting failed in the Cherubin setting~\cite{cherubin2022online}. We attribute this to three possible methodological differences. First, \textit{exit-to-entry repositioning}: their classifier was trained on exit traces but tested on entry traces, introducing feature distortion. Second, \textit{label granularity}: domain labels are harder to classify than page labels as domain labels may correspond to a large number of pages. Third, \textit{training set size}: in their final results, a given class could have as few as 10 traces in training.
We cannot determine which factor is dominant as their data cannot be made available due to privacy concerns.

We can turn to Jansen et al.~\cite{jansen2023repositioning} to isolate the cause. Even after simulating trace transformation to resolve the exit-to-entry repositioning issue, they reported a low median $F_1$ score of 0.35. This suggests that repositioning was not the primary factor, pointing instead to label granularity or training set size. These two factors compound: any specific page might form only a small portion of the training data, and the classifier would not be able to train on its traffic characteristics.
With a sufficient training set and page labels, we demonstrated that a classifier trained and tested on synthetic monitored traces succeeds in a real Tor open world, even with cross-network vantage mismatch. Compared to Cherubin et al.'s final experiment, the only difference lies in the training strategy, which is up to the attacker.

\subsection{Why Do We Use Synthetic Monitored Traces?}

Cherubin et al.~\cite{cherubin2022online} provide compelling arguments for testing with synthetic monitored traces despite the goal of realism: (i) it reduces the privacy compromise of real users (ii) exit-collected real monitored traces do not have page-labeled ground truth. We further argue that synthetic monitored traces are sufficient to demonstrate real threat with an analysis of the differences between how a user and a crawler visits web pages. Both actions generate a GET request on the network, after which the browser handles resource loading.

While trace differences could theoretically arise from cookies, Tor Browser deletes persistent cookies upon exit, neutralizing long-term tracking. Similarly, personalization based on geography or IP address depends on the exit node, not the client. Furthermore, Tor Browser standardizes browser and OS settings to mitigate browser fingerprinting~\cite{tor-browser-fingerprinting}, effectively homogenizing the client-side environment. Finally, while unpredictable user behavior (e.g., interrupting a load or opening parallel tabs) is not captured in synthetic traces, the demonstrated vulnerability of clean, page-load-aligned traces is sufficient to show that safeguarding anonymity on Tor requires stronger defenses.

One limitation of this and prior work is that synthetic traces do not capture dynamic web pages, which dominate the most popular websites. An ameliorating factor on Tor Browser is that without cookies, users would often need to login to such websites through a static page, meaning an attacker can still see the user is visiting that website by monitoring the login page. Moreover, the most sensitive pages are likely not the most popular; many sensitive resources (such as protester guides and whistleblower platforms) are static pages and their use is captured well with synthetic traces.

\section{Conclusion and Future Work}
\label{sec:conclusion}

This work re-evaluates the threat of WF using a privacy-preserving methodology that allowed us to collect real non-monitored traces at a guard relay. We demonstrate that WF attacks remain highly effective in the real Tor open world when trained on precise webpage labels, with the DF attack achieving strong results even under cross-network conditions, with or without trace isolation capabilities from the guard vantage point. We further demonstrate that the best WF attacks survive small training set sizes (with a minimal loss in recall at 70 traces per page), benefit from larger real non-monitored sets, and remain fairly robust to concept drift (suffering an $F_1$ score drop of 0.238 over six months). As we used synthetic monitored traces like previous works, the next step would be to evaluate whether real, well-labeled monitored traces are more difficult to classify than synthetic traces, and whether this difference can be surmounted. Finally, we show that while Tor's Conflux protocol weakens the guard as a WF attacker by splitting traffic, it is not a silver bullet; a guard with a latency advantage can exploit the default LowRTT scheduling to capture the feature-rich start of the connection and recover attack performance. Future work can address this vulnerability by designing Conflux scheduling algorithms that mitigate latency bias.

\cleardoublepage
\appendix

\section*{Ethical Considerations}

\subsection*{Stakeholders}
The primary stakeholders in this study are:

\vspace{-5.5pt}
\paragraph{Tor Users (Data Subjects).} Real-world Tor users whose encrypted traffic was routed through our guard relay, particularly those in censorship-heavy regions who rely on Tor for safety.

\vspace{-5.5pt}
\paragraph{The Tor Community.} The network of volunteers and developers who rely on the integrity of the network design.

\vspace{-5.5pt}
\paragraph{The Research Community.} Researchers who benefit from high-quality, realistic Tor datasets to benchmark and improve privacy defenses.

\vspace{-5.5pt}
\paragraph{Adversaries (Indirect Stakeholders).} Entities such as censoring regimes, ISPs, or malicious actors, who may seek to exploit the weaknesses exposed in this research.

\subsection*{Impacts}
We analyzed the impact of demonstrating WF attack effectiveness in the real world---specifically by gathering real non-monitored traces combined with synthetic monitored traces---through the lens of ethical principles: \textit{Respect for Persons} (Privacy), \textit{Beneficence} (Network Security), and \textit{Non-Maleficence} (Do No Harm).

\vspace{-5.5pt}
\paragraph{Harms to Privacy (Tor Users).} The primary ethical risk is the potential violation of user privacy via metadata collection. Collecting real traffic carries a theoretical risk of linkability: if a collected trace is sufficiently unique, it could expose what a user is doing on Tor, which violates their privacy.

\vspace{-5.5pt}
\paragraph{Risk of Misuse (Dual-Use).} By demonstrating the potency of website fingerprinting (WF) attacks in the wild, there is a risk that malicious actors may be motivated to de-anonymize real users by adopting a similar methodology without our safeguards.

\vspace{-5.5pt}
\paragraph{Benefits (Network Security).} The study provides insight into the realistic threat of WF attacks. By demonstrating that attacks remain effective in the open world, this work serves to inform the Tor and research communities. It motivates the development of stronger defenses, ultimately benefiting the privacy of all Tor users.

\subsection*{Mitigations}
We designed our framework to strictly adhere to data minimization principles, ensuring that collecting real-world traffic poses no privacy risk while providing necessary data quality. Real-world traffic is utilized solely to provide unlabeled, non-monitored traces for classification. Below, we detail our handling, filtering, and sanitization procedures.

\vspace{-5.5pt}
\paragraph{Trace Reconstruction \& Ephemeral Identifiers.} We recorded only the following metadata for each non-monitored Tor cell: channel ID, circuit ID, direction, and timestamp. We utilized \textit{ephemeral-only}, relay-local identifiers (channel/circuit IDs) that cannot be linked back to any user (see Section~\ref{sec:circuits_and_guards}). We did not record any cell payloads.

\vspace{-5.5pt}
\paragraph{In-Memory IP Filtering.} While the guard relay must inevitably process client IP addresses to route cells, we strictly adhered to a policy of never writing real user IP addresses to disk. We used in-memory filtering to check the IP address of a live channel against a pre-defined list of our controlled crawlers, separating monitored from non-monitored traffic immediately (Section~\ref{sec:guard-mod}). We further utilized Tor's built-in authentication scheme to exclude relay-to-relay traffic (Section~\ref{sec:non-mon-traffic}).

\vspace{-5.5pt}
\paragraph{Trace Sanitization.} Ephemeral identifiers (channel/circuit IDs) were exclusively used to demultiplex the stream of cells into traces; they were removed from the final dataset. The final non-monitored dataset consists solely of sequences of directions and timestamps. Furthermore, all non-monitored traces were shuffled and time-normalized (start time reset to zero) to prevent temporal linkage or correlation with external server logs.

\vspace{-5.5pt}
\paragraph{Dataset Access Control.} Although we believe our pipeline ensures safe, anonymous release of Tor traffic, in the interest of prudence, we limit access to the research community. This restriction will remain in effect for one year following our initial data release to minimize time sensitivity.
\subsection*{Decisions}

\paragraph{Methodological Necessity.} We chose to collect real background traffic because synthetic non-monitored data fails to capture the complexity of the real open world~\cite{cherubin2022online}. We concluded that the security benefits of accurately benchmarking these threats outweigh the privacy risks, provided those risks are minimized through the strict methodology described above.

\vspace{-5.5pt}
\paragraph{Balancing Publication vs. Misuse.} We acknowledge the dual-use nature of this work. However, we decided that suppressing these findings would be more harmful to the Tor ecosystem than publishing them. ``Security by obscurity'' leaves users vulnerable to resourceful adversaries (such as state-level actors) who may already be capable of these attacks. By publicizing the realistic nature of WF attacks, we empower the Tor Project and community to engineer concrete defenses.

\clearpage

\section*{Open Science}

To foster reproducibility and support future research, we have released our code and datasets to the public~\cite{artifacts_2026}. Specifically, the released datasets are sanitized and include all high-quality traces that we have collected throughout this work and can be used for future research. They are permanently archived and available via the Open Science Framework (OSF) at the following repository:  

\vspace{5pt}
\noindent
URL: \url{https://osf.io/9m8ea/} \\
DOI: \url{https://doi.org/10.17605/OSF.IO/9M8EA}

\vspace{5pt}
Please note that while the synthetic monitored datasets are openly accessible under a standard non-commercial license, access to the real-world non-monitored datasets is temporarily restricted. Consistent with the privacy safeguards detailed in \textit{Ethical Considerations} appendix, access to this restricted data will only be granted to the research community under a Data Use Agreement for one year following the initial release.


\begin{thebibliography}{10}

\bibitem{alsabah2013path}
Mashael AlSabah, Kevin Bauer, Tariq Elahi, and Ian Goldberg.
\newblock {The Path Less Travelled: Overcoming Tor's Bottlenecks with Traffic Splitting}.
\newblock In {\em Proceedings of the 13th Privacy Enhancing Technologies Symposium (PETs)}, pages 143--163. Springer, 2013.

\bibitem{Bhat_2019}
Sanjit Bhat, David Lu, Albert Kwon, and Srinivas Devadas.
\newblock {Var-CNN: A Data-Efficient Website Fingerprinting Attack Based on Deep Learning}.
\newblock {\em Proceedings on Privacy Enhancing Technologies (PoPETs)}, 2019(4):292--310, 2019.

\bibitem{breslau1999web}
Lee Breslau, Pei Cao, Li~Fan, Graham Phillips, and Scott Shenker.
\newblock {Web Caching and Zipf-like Distributions: Evidence and Implications}.
\newblock In {\em Proceedings of the IEEE INFOCOM}, volume~1, pages 126--134, 1999.

\bibitem{brown2001interval}
Lawrence~D Brown, T~Tony Cai, and Anirban DasGupta.
\newblock {Interval Estimation for a Binomial Proportion}.
\newblock {\em Statistical Science}, 16(2):101--133, 2001.

\bibitem{cassel2022omnicrawl}
Darion Cassel, Su-Chin Lin, Alessio Buraggina, William Wang, Andrew Zhang, Lujo Bauer, Hsu-Chun Hsiao, Limin Jia, and Timothy Libert.
\newblock {OmniCrawl: Comprehensive Measurement of Web Tracking with Real Desktop and Mobile Browsers}.
\newblock {\em Proceedings on Privacy Enhancing Technologies (PoPETs)}, 2022(1):227--252, 2022.

\bibitem{cherubin2022online}
Giovanni Cherubin, Rob Jansen, and Carmela Troncoso.
\newblock {Online Website Fingerprinting: Evaluating Website Fingerprinting Attacks on Tor in the Real World}.
\newblock In {\em Proceedings of the 31st USENIX Security Symposium (USENIX Security 22)}, pages 753--770, 2022.

\bibitem{sayrafi_2018}
{Cloudflare}.
\newblock {Introducing the Cloudflare Onion Service}.
\newblock \url{https://blog.cloudflare.com/cloudflare-onion-service/}.
\newblock Accessed: 2026-01-16.

\bibitem{cloudflare_docs_2023}
{Cloudflare}.
\newblock {Onion Routing and Tor Support}.
\newblock \url{https://developers.cloudflare.com/network/onion-routing/}.
\newblock Accessed: 2026-01-16.

\bibitem{deng2024robust}
Xinhao Deng, Qi~Li, and Ke~Xu.
\newblock {Robust and Reliable Early-Stage Website Fingerprinting Attacks via Spatial-Temporal Distribution Analysis}.
\newblock In {\em Proceedings of the 2024 ACM SIGSAC Conference on Computer and Communications Security (CCS)}, pages 1997--2011, 2024.

\bibitem{robust_multitab_wf}
Xinhao Deng, Qilei Yin, Zhuotao Liu, Xiyuan Zhao, Qi~Li, Mingwei Xu, Ke~Xu, and Jianping Wu.
\newblock {Robust Multi-Tab Website Fingerprinting Attacks in the Wild}.
\newblock In {\em Proceedings of the IEEE Symposium on Security and Privacy (S\&P)}, pages 1005--1022, 2023.

\bibitem{deng2026towards}
Xinhao Deng, Xiyuan Zhao, Qilei Yin, Zhuotao Liu, Qi~Li, Mingwei Xu, Ke~Xu, and Jianping Wu.
\newblock {Towards robust multi-tab website fingerprinting}.
\newblock {\em IEEE Transactions on Networking}, 2026.
\newblock Early Access.

\bibitem{dingledine2004tor}
Roger Dingledine, Nick Mathewson, and Paul Syverson.
\newblock {Tor: The Second-Generation Onion Router}.
\newblock In {\em Proceedings of the 13th USENIX Security Symposium (USENIX Security 04)}, San Diego, CA, 2004.

\bibitem{gong2020zero}
Jiajun Gong and Tao Wang.
\newblock {Zero-Delay Lightweight Defenses Against Website Fingerprinting}.
\newblock In {\em Proceedings of the 29th USENIX Security Symposium (USENIX Security 20)}, pages 717--734, 2020.

\bibitem{guan2021bapm}
Zhong Guan, Gang Xiong, Gaopeng Gou, Zhen Li, Mingxin Cui, and Chang Liu.
\newblock {BAPM: Block Attention Profiling Model for Multi-Tab Website Fingerprinting Attacks on Tor}.
\newblock In {\em Proceedings of the 37th Annual Computer Security Applications Conference (ACSAC)}, pages 248--259, 2021.

\bibitem{hayes2016k}
Jamie Hayes and George Danezis.
\newblock {k-Fingerprinting: A Robust Scalable Website Fingerprinting Technique}.
\newblock In {\em Proceedings of the 25th USENIX Security Symposium (USENIX Security 16)}, pages 1187--1203, 2016.

\bibitem{jansen2023repositioning}
Rob Jansen, Ryan Wails, and Aaron Johnson.
\newblock {Repositioning Real-World Website Fingerprinting on Tor}.
\newblock In {\em Proceedings of the 23rd Workshop on Privacy in the Electronic Society (WPES)}, pages 124--140, 2023.

\bibitem{jin2023transformer}
Zhaoxin Jin, Tianbo Lu, Shuang Luo, and Jiaze Shang.
\newblock {Transformer-Based Model for Multi-Tab Website Fingerprinting Attack}.
\newblock In {\em Proceedings of the 2023 ACM SIGSAC Conference on Computer and Communications Security (CCS)}, pages 1050--1064, 2023.

\bibitem{juarez2014critical}
Marc Juarez, Sadia Afroz, Gunes Acar, Claudia Diaz, and Rachel Greenstadt.
\newblock {A Critical Evaluation of Website Fingerprinting Attacks}.
\newblock In {\em Proceedings of the 2014 ACM SIGSAC Conference on Computer and Communications Security (CCS)}, pages 263--274, 2014.

\bibitem{krumnow2022how}
Benjamin Krumnow, Hugo Jonker, and Stefan Karsch.
\newblock {How Gullible Are Web Measurement Tools? A Case Study Analysing and Strengthening OpenWPM's Reliability}.
\newblock In {\em Proceedings of the 18th International Conference on Emerging Networking EXperiments and Technologies (CoNEXT '22)}, pages 171--186, 2022.

\bibitem{LePochat2019}
Victor Le~Pochat, Tom Van~Goethem, Samaneh Tajalizadehkhoob, Maciej Korczynski, and Wouter Joosen.
\newblock {Tranco: A Research-Oriented Top Sites Ranking Hardened Against Manipulation}.
\newblock In {\em Proceedings of the Network and Distributed System Security Symposium (NDSS)}. Internet Society, 2019.

\bibitem{mani2018understanding}
Akshaya Mani, T~Wilson-Brown, Rob Jansen, Aaron Johnson, and Micah Sherr.
\newblock {Understanding Tor Usage with Privacy-Preserving Measurement}.
\newblock In {\em Proceedings of the Internet Measurement Conference (IMC)}, pages 175--187, 2018.

\bibitem{rfc7838}
Mark Nottingham, Patrick McManus, and Julian Reschke.
\newblock {HTTP Alternative Services}.
\newblock RFC 7838, Apr 2016.

\bibitem{panchenko2016website}
Andriy Panchenko, Fabian Lanze, Andreas Zinnen, Martin Henze, Jan Pennekamp, Klaus Wehrle, and Thomas Engel.
\newblock {Website Fingerprinting at Internet Scale}.
\newblock In {\em Proceedings of the Network and Distributed System Security Symposium (NDSS)}. Internet Society, 2016.

\bibitem{Rahman_2020}
Mohammad~Saidur Rahman, Payap Sirinam, Nate Mathews, Kantha~Girish Gangadhara, and Matthew Wright.
\newblock {Tik-Tok: The Utility of Packet Timing in Website Fingerprinting Attacks}.
\newblock {\em Proceedings on Privacy Enhancing Technologies (PoPETs)}, 2020(3):5--24, 2020.

\bibitem{Rimmer_2018}
Vera Rimmer, Davy Preuveneers, Marc Juarez, Tom Van~Goethem, and Wouter Joosen.
\newblock {Automated Website Fingerprinting through Deep Learning}.
\newblock In {\em Proceedings of the Network and Distributed System Security Symposium (NDSS)}. Internet Society, 2018.

\bibitem{artifacts_2026}
Mohammadhamed Shadbeh, Khashayar Khajavi, and Tao Wang.
\newblock {Reality Check for Tor Website Fingerprinting in the Open World}.
\newblock \url{https://doi.org/10.17605/OSF.IO/9M8EA}, 2026.
\newblock OSF Repository. DOI: 10.17605/OSF.IO/9M8EA.

\bibitem{shen2023subverting}
Meng Shen, Kexin Ji, Zhenbo Gao, Qi~Li, Liehuang Zhu, and Ke~Xu.
\newblock {Subverting Website Fingerprinting Defenses with Robust Traffic Representation}.
\newblock In {\em Proceedings of the 32nd USENIX Security Symposium (USENIX Security 23)}, pages 607--624, 2023.

\bibitem{sirinam2018deep}
Payap Sirinam, Mohsen Imani, Marc Juarez, and Matthew Wright.
\newblock {Deep Fingerprinting: Undermining Website Fingerprinting Defenses with Deep Learning}.
\newblock In {\em Proceedings of the 2018 ACM SIGSAC Conference on Computer and Communications Security (CCS)}, pages 1928--1943, 2018.

\bibitem{sirinam2019triplet}
Payap Sirinam, Nate Mathews, Mohammad~Saidur Rahman, and Matthew Wright.
\newblock {Triplet Fingerprinting: More Practical and Portable Website Fingerprinting with N-Shot Learning}.
\newblock In {\em Proceedings of the 2019 ACM SIGSAC Conference on Computer and Communications Security (CCS)}, pages 1131--1148, 2019.

\bibitem{tor-browser-fingerprinting}
{Tor Project}.
\newblock {How Tor Browser Protects You Against Browser Fingerprinting}.
\newblock \url{https://support.torproject.org/tor-browser/features/fingerprinting-protections}.
\newblock Accessed: 2026-01-16.

\bibitem{perry2013critique}
{Tor Project}.
\newblock {Tor Blog - A Critique of Website Traffic Fingerprinting Attacks}.
\newblock \url{https://blog.torproject.org/critique-website-traffic-fingerprinting-attacks/}.
\newblock Accessed: 2026-01-16.

\bibitem{tor_blog_guard}
{Tor Project}.
\newblock {Tor Blog - The Lifecycle of a New Relay}.
\newblock \url{https://blog.torproject.org/lifecycle-of-a-new-relay/}.
\newblock Accessed: 2026-01-16.

\bibitem{stream_isolation}
{Tor Project}.
\newblock {Tor Design Proposals \#171 - Separate Streams Across Circuits by Connection Metadata}.
\newblock \url{https://spec.torproject.org/proposals/171-separate-streams.html}.
\newblock Accessed: 2026-01-16.

\bibitem{tor-prop329}
{Tor Project}.
\newblock {Tor Design Proposals \#329 - Overcoming Tor's Bottlenecks with Traffic Splitting}.
\newblock \url{https://spec.torproject.org/proposals/329-traffic-splitting.html}.
\newblock Accessed: 2026-01-16.

\bibitem{start_tor_browser_gitlab}
{Tor Project}.
\newblock {Tor Project GitLab - Using a System-Installed Tor Process with Tor Browser}.
\newblock \url{https://gitlab.torproject.org/tpo/applications/tor-browser-build/-/blob/a12487a8d73b1a131d0b7402e3d4eb52c58f6cf4/projects/browser/RelativeLink/start-browser#L341}.
\newblock Accessed: 2026-01-16.

\bibitem{tor-spec}
{Tor Project}.
\newblock {Tor Protocol Specification}.
\newblock \url{https://spec.torproject.org/tor-spec/}.
\newblock Accessed: 2026-01-16.

\bibitem{w3techs-cloudflare-reverse-proxy-2026}
{W3Techs}.
\newblock {Usage Statistics and Market Share of Cloudflare}.
\newblock \url{https://w3techs.com/technologies/details/cn-cloudflare}.
\newblock Accessed: 2026-01-16.

\bibitem{wang2020high}
Tao Wang.
\newblock {High Precision Open-World Website Fingerprinting}.
\newblock In {\em Proceedings of the IEEE Symposium on Security and Privacy (S\&P)}, pages 152--167. IEEE, 2020.

\bibitem{wang2014effective}
Tao Wang, Xiang Cai, Rishab Nithyanand, Rob Johnson, and Ian Goldberg.
\newblock {Effective Attacks and Provable Defenses for Website Fingerprinting}.
\newblock In {\em Proceedings of the 23rd USENIX Security Symposium (USENIX Security 14)}, pages 143--157, 2014.

\bibitem{wang2013improved}
Tao Wang and Ian Goldberg.
\newblock {Improved Website Fingerprinting on Tor}.
\newblock In {\em Proceedings of the 12th ACM Workshop on Privacy in the Electronic Society (WPES)}, pages 201--212, 2013.

\bibitem{wang2016realistically}
Tao Wang and Ian Goldberg.
\newblock {On Realistically Attacking Tor with Website Fingerprinting}.
\newblock {\em Proceedings on Privacy Enhancing Technologies (PoPETs)}, 2016(4):21--36, 2016.

\bibitem{wang2022snwf}
Yanbin Wang, Haitao Xu, Zhenhao Guo, Zhan Qin, and Kui Ren.
\newblock {SnWF: Website Fingerprinting Attack by Ensembling the Snapshot of Deep Learning}.
\newblock {\em IEEE Transactions on Information Forensics and Security}, 17:1214--1226, 2022.

\bibitem{xu2018multi}
Yixiao Xu, Tao Wang, Qi~Li, Qingyuan Gong, Yang Chen, and Yong Jiang.
\newblock {A Multi-Tab Website Fingerprinting Attack}.
\newblock In {\em Proceedings of the 34th Annual Computer Security Applications Conference (ACSAC)}, pages 327--341, 2018.

\bibitem{automated_multitab_wf}
Qilei Yin, Zhuotao Liu, Qi~Li, Tao Wang, Qian Wang, Chao Shen, and Yixiao Xu.
\newblock {An Automated Multi-Tab Website Fingerprinting Attack}.
\newblock {\em IEEE Transactions on Dependable and Secure Computing}, 19(6):3656--3670, 2022.

\end{thebibliography}



\section{Models’ Hyperparameters}
\label{app:hyperparams}

\begin{table*}[htbp] 
\centering
\caption{Hyperparameter settings for the Website Fingerprinting models.}
\label{tab:hyperparameters}

\small

\renewcommand{\arraystretch}{1.5} 

\begin{tabularx}{\textwidth}{@{} l *{5}{>{\raggedright\arraybackslash}X} @{}}
\toprule
\textbf{Parameter} & \textbf{$k$-FP} & \textbf{DF} & \textbf{Tik-Tok} & \textbf{RF} & \textbf{Holmes} \\ 
\midrule
\textbf{Model Type} & Random Forest & 1D CNN & 1D CNN & 2D CNN & Dual-Branch CNN \\ 
\addlinespace
\textbf{Batch Size} & N/A & 128 & 32 & 200 & 200 / 256 \\ 
\addlinespace
\textbf{Epochs} & N/A & 30 & 30 (max) & 30 & 30 \\ 
\addlinespace
\textbf{Optimizer} & N/A & Adamax & Adamax & Adam & Adam / AdamW \\ 
\addlinespace

\textbf{Learning Rate} & N/A & 0.002 & 0.002 & $5\cdot 10^{-4} \times$ \newline $0.2^{(\frac{\text{epoch}}{30})}$ & 0.0005 \\ 
\addlinespace
\textbf{Loss Function} & N/A & CrossEntropy & CrossEntropy & CrossEntropy & CrossEntropy / \newline SupConLoss \\ 
\addlinespace

\textbf{Input Dims} & Feature Vector: \newline 175 (max) & Length: 5,000 & Length: 5,000 & TAM length: 1800 & Temporal: 1000 \newline TAF: 2000 \\ 
\addlinespace
\textbf{Other Params} & $n_{est}=1,000$ \newline feat. importances \newline used & $\beta_1=0.9$ \newline $\beta_2=0.999$ \newline $\epsilon=10^{-8}$ & $\beta_1=0.9, \beta_2=0.999$ \newline $\epsilon=10^{-8}$ \newline EarlyStop: val\_loss \newline patience=6 & Maximum load\_time: \newline Max trace load \newline (Exp. dependent) & Best F1-score \newline metric for \newline epochs used \\ 
\bottomrule
\end{tabularx}
\end{table*}
 
To support the reproducibility of our experiments, Table~\ref{tab:hyperparameters} details the architectural and training configurations for the Website Fingerprinting models evaluated. The table aggregates hyperparameters for the feature-based $k$-FP model as well as the deep learning classifiers (DF, Tik-Tok, RF, and Holmes).

We report the specific optimizer settings, batch sizes, and input dimensions used to generate our results. While we largely maintained the default settings of the original models, we adjusted the maximum load time for the RF model’s TAM features. To optimize performance, this load time was set dynamically based on the highest trace load time in each experiment (ranging from 42--47 seconds), with the exception of baseline experiments where the default of 80 seconds was retained.

\subsection{Impact of RF Hyperparameters}
\label{app:rf-hyper}
The performance of the RF model degrades significantly when there is a substantial latency discrepancy between the training and testing data. This affected the performance of this model in our experiments. In our experimental setup, we observed a latency difference of approximately 150~ms between the Round Trip Time (RTT) of the AU client (training set) and the CA client (testing set).

The RF model featurizes traces using a Traffic Aggregation Matrix (TAM). The duration of a single time slot within this matrix is determined by the maximum load time ($T_{max}$) divided by the TAM length ($N$), such that $t_{slot} = T_{max} / N$. By default, standard implementations often use $T_{max}=80s$ and $N=1800$, resulting in a slot duration of $\approx$44~ms. To investigate the relationship between slot granularity and latency shifts, we varied the TAM length to achieve different slot durations, while normalizing $T_{max}$ to the longest trace in each experiment (45s in this scenario).

\begin{figure}[t!]
    \centering
    \includegraphics[width=1\linewidth]{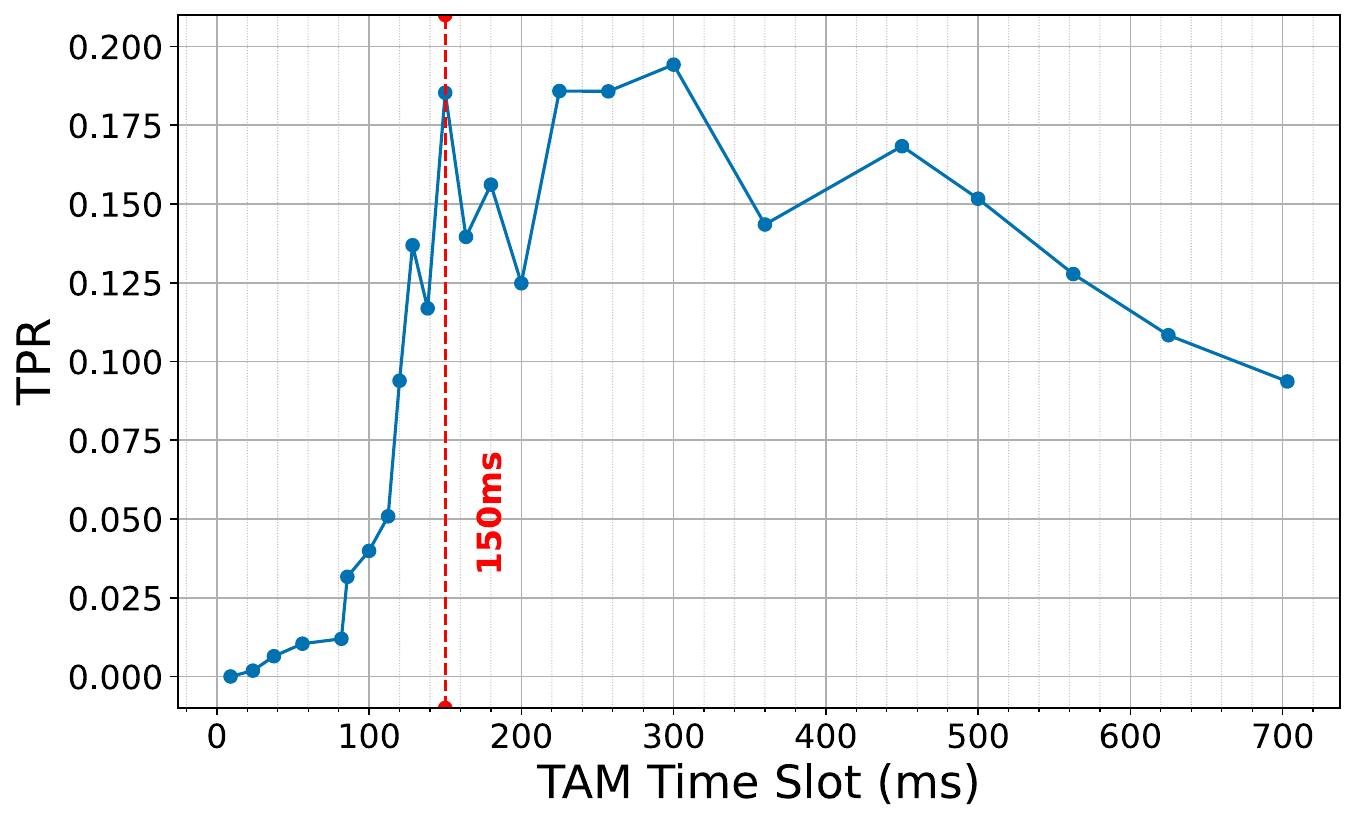}
    \caption{Impact of Time Slot duration on RF performance (threshold tuned for a fixed FPR of 0.5\%). The red dashed line marks the 150~ms latency difference between the training (AU) and testing (CA) regions. Performance improves significantly once the slot size exceeds this latency threshold.}
    \label{fig:rf-params-slots}
\end{figure}

Figure~\ref{fig:rf-params-slots} illustrates the model’s TPR at a fixed FPR of 0.5\% across varying time slot durations. The model exhibits poor performance with fine-grained time slots; for example, at the default $\approx$44~ms slot size, the recall is negligible~($\approx$0.01). This suggests that small bins are highly sensitive to the temporal misalignment caused by the regional latency difference.

However, a sharp increase in performance is observed as the time slot duration approaches the 150~ms latency delta. The TPR rises to $\approx$0.18 at the 150~ms mark, stabilizing shortly after. The maximum TPR of $\approx$0.19 is achieved at a slot duration of 300~ms. This indicates that larger time slots can smooth out the jitter introduced by the network conditions. However, RF still fails to achieve performance on par with other models such as DF, which achieves a TPR of $\approx$0.94 under similar conditions.

\end{document}